\documentclass[aps,twocolumn,reprint,superscriptaddress,amsmath,amssymb]{revtex4-1}

\usepackage[dvipdfmx]{graphicx}
\usepackage{dcolumn}
\usepackage{bm}
\usepackage{color}
\usepackage{textcomp}

\begin{document}


\title{Magnetic Structural Unit with Convex Geometry: a Building Block Hosting an Exchange-striction-driven Magnetoelectric Coupling}

\author{Kenta Kimura}
\email[]{kentakimura@edu.k.u-tokyo.ac.jp}
\affiliation{Department of Advanced Materials Science, University of Tokyo, Kashiwa, Chiba 277-8561, Japan}
\author{Yasuyuki Kato}
\affiliation{Department of Applied Physics, The University of Tokyo, Hongo, 7-3-1, Bunkyo, Tokyo 113-8656, Japan}
\author{Kunihiko Yamauchi}
\affiliation{ISIR-SANKEN, Osaka University, Ibaraki, Osaka 567-0047, Japan}
\author{Atsushi Miyake}
\affiliation{Institute for Solid State Physics, The University of Tokyo, Kashiwa, Chiba 277-8581, Japan}
\author{Masashi Tokunaga}
\affiliation{Institute for Solid State Physics, The University of Tokyo, Kashiwa, Chiba 277-8581, Japan}
\author{Akira Matsuo}
\affiliation{Institute for Solid State Physics, The University of Tokyo, Kashiwa, Chiba 277-8581, Japan}
\author{Koichi Kindo}
\affiliation{Institute for Solid State Physics, The University of Tokyo, Kashiwa, Chiba 277-8581, Japan}
\author{Mitsuru Akaki}
\affiliation{Center for Advanced High Magnetic Field Science, Graduate School of Science, Osaka University,
Toyonaka, Osaka 560-0043, Japan}
\author{Masayuki Hagiwara}
\affiliation{Center for Advanced High Magnetic Field Science, Graduate School of Science, Osaka University,
Toyonaka, Osaka 560-0043, Japan}
\author{Shojiro Kimura}
\affiliation{Institute for Materials Research, Tohoku University, Katahira 2-1-1, Sendai 980-8577, Japan}
\author{Masayuki Toyoda}
\affiliation{Department of Physics, Tokyo Institute of Technology, Meguro-ku, Tokyo 152-8550, Japan}
\author{Yukitoshi Motome}
\affiliation{Department of Applied Physics, The University of Tokyo, Hongo, 7-3-1, Bunkyo, Tokyo 113-8656, Japan}
\author{Tsuyoshi Kimura}
\affiliation{Department of Advanced Materials Science, University of Tokyo, Kashiwa, Chiba 277-8561, Japan}

\email[*]{kentakimura@edu.k.u-tokyo.ac.jp}
\altaffiliation{aaa}


\date{\today}

\begin{abstract}
We perform a combined experimental and theoretical study of a magnetic-field ($B$) induced evolution of magnetic and ferroelectric properties in an antiferromagnetic material Pb(TiO)Cu$_4$(PO$_4$)$_4$, whose structure is characterized by a staggered array of Cu$_4$O$_{12}$ magnetic units with convex geometry known as square cupola. Our experiments show a $B$-induced phase transition from a previously reported low-$B$ linear magnetoelectric phase to a new high-$B$ magnetoelectric phase, which accompanies a 90$^\circ$ flop of electric polarization and gigantic magnetodielectric effect. Moreover, we observe a $B$-induced sign reversal of ferroelectric polarization in the high-$B$ phase.  Our model and first-principles calculations reveal that the observed complex magnetoelectric behavior is well explained in terms of a $B$-dependent electric polarization generated in each  Cu$_4$O$_{12}$ unit by the so-called exchange striction mechanism. The present study demonstrates that the materials design based on the magnetic structural unit with convex geometry deserves to be explored for developing strong magnetoelectric couplings.

\end{abstract}



\maketitle


\section{INTORODUCTION}
Magnetoelectric multiferroics, in which magnetic and ferroelectric orders coexist, are important class of materials because their unique and strong magnetoelectric couplings provide numerous potential applications such as novel magneto-optical devices and antiferromagnetic spintronics devices \cite{Astrov1960,Dubovik1990,Kimura2003,Fiebig2005,cheong2007multiferroics,arima2008magneto,Spaldin2008,mundy2016atomically,baltz2018antiferromagnetic}. Recently, designing magnetoelectric multiferroic materials based on structural units such as specific molecules or transition metal ion clusters has been extensively studied. Experimentally, many molecular-based multiferroic materials have been found particularly in metal-organic hybrid systems \cite{jain2009multiferroic, xu2011coexistence, pato2013coexistence}. In most of them, their magnetic order is provided by magnetic moments of transition metal ions while their ferroelectric order is associated with an order-disorder transition of organic molecular units. Because of this different origin of magnetic and ferroelectric orders, however, their magnetoelectric coupling is generally weak \cite{ramesh2009materials}, and hence a drastic response of an electric polarization (magnetization) to an external magnetic (electric) field has been merely observed.

A promising way to enhance magnetoelectric couplings has been already known from extensive studies on magnetoelectric multiferroic inorganic oxides, where magnetic order itself generates an electric polarization \cite{Kimura2003}. Three types of mechanisms for an induced polarization are well established: the spin current mechanism \cite{Katsura2005} (or the inverse Dzyaloshinskii-Moriya (DM) interaction mechanism \cite{Sergienko2006}), the metal-ligand $d$-$p$ hybridization mechanism \cite{Arima2007}, and the exchange striction mechanism \cite{Sergienko2006ferroelectricity}. The former two mechanisms are associated with a weak relativistic spin-orbit interaction, which usually yields a small polarization. On the other hand, the exchange striction mechanism, not involving the spin-orbit interaction, potentially generates a much larger polarization, as observed in perovskite manganites such as pressurized $R$MnO$_3$ ($R$ = Tb, Dy, and Gd) \cite{aoyama2014giant,aoyama2015multiferroicity,terada2016magnetic}. 
Therefore, the use of magnetic structural units, where the exchange striction mechanism is active, is a key to designing a material with a strong magnetoelectric coupling. Theoretically, a system consisting of magnetic trimer molecules was proposed to exhibit an exchange-striction-driven strong magnetoelectric coupling \cite{Delaney2009}. However, this proposal has not been confirmed experimentally.

\begin{figure}[t]
\includegraphics[width=8.6cm]{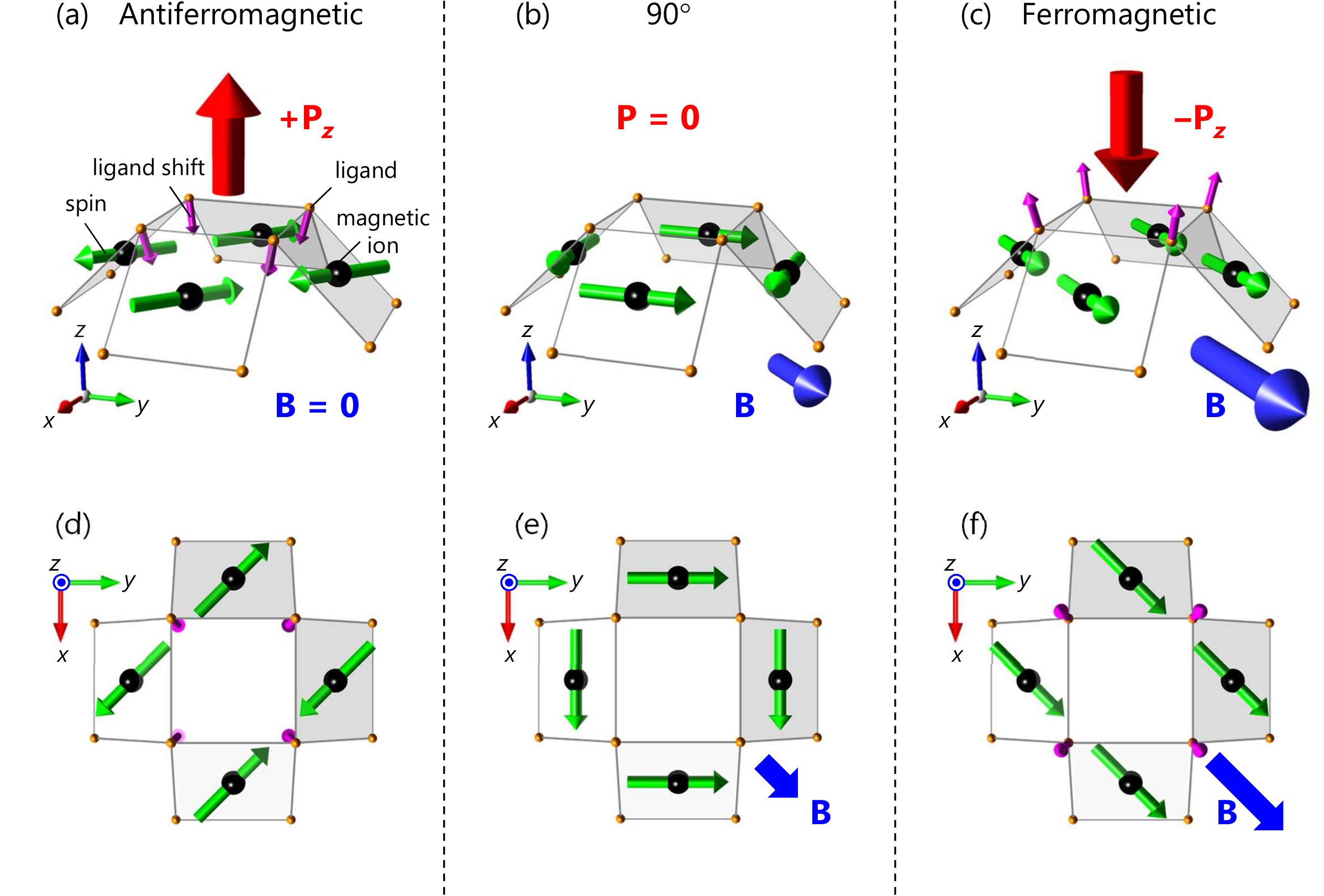}
\caption{
Proposal for a magnetically-controllable electric polarization via the exchange striction using a convex square cupola spin cluster. The black and orange balls represent the magnetic ion and ligand (anion), respectively. The green arrows denote a spin. The pink arrows indicate a ligand shift due to an exchange striction effect. This shift induces an electric dipole along the opposite direction.  The electric polarization ({\bf P}) appears parallel and antiparallel to the convex direction for 
(a) antiferromagnetic (AFM) and (c) ferromagnetic  (FM) cases, respectively, while no  {\bf P} appears for (b) the 90$^\circ$ arrangement case. 
Therefore, when the spin arrangement is changed from AFM to FM by a magnetic field ({\bf B}),  the direction of {\bf P} should be reversed. 
For clarity, top view illustrations of the panels (a), (b), and (c) are shown in (d), (e), and (f), respectively.
\label{design}
}
\end{figure}

\begin{figure}[t]
\includegraphics[width=7cm]{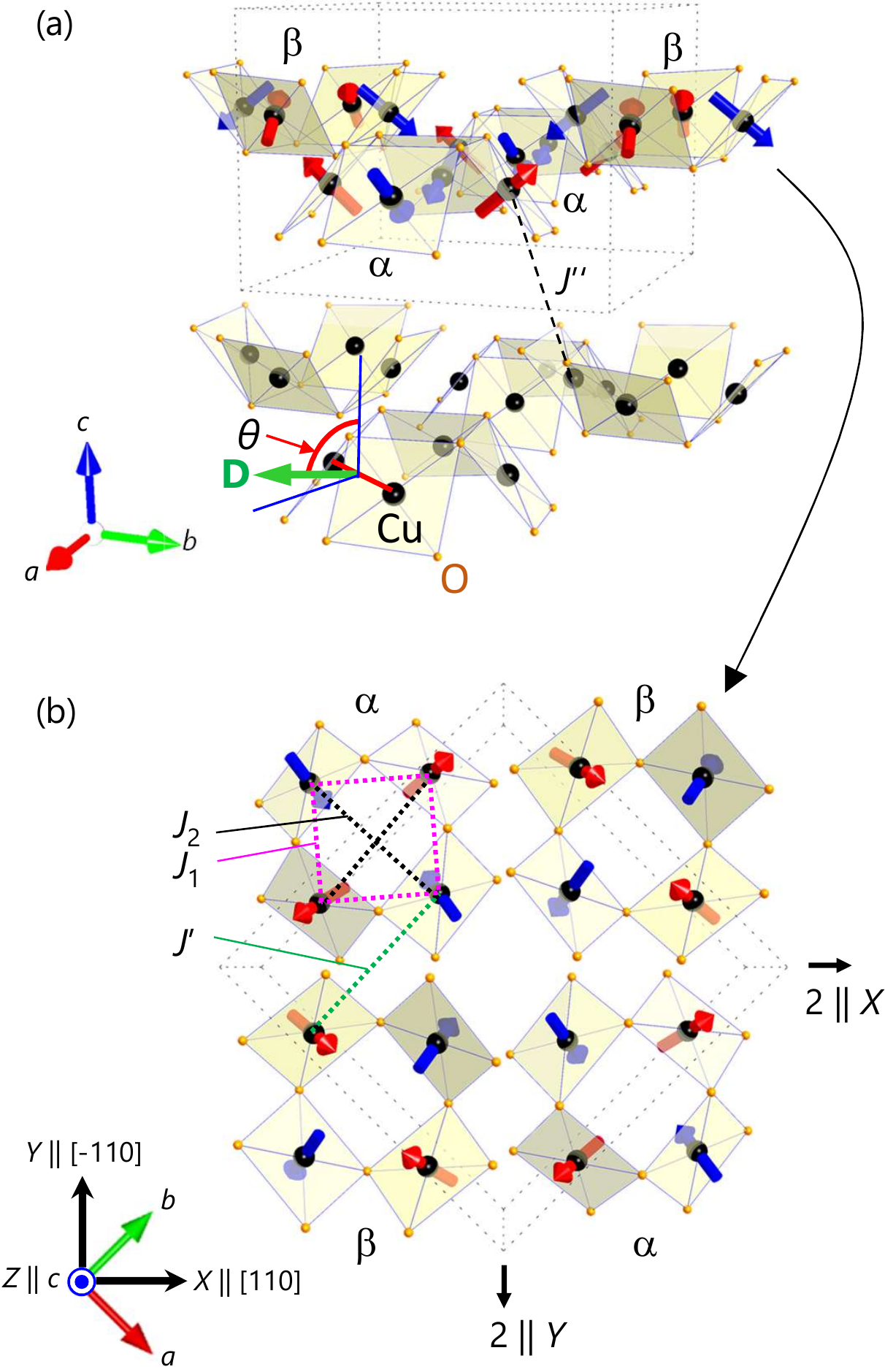}
\caption{
Crystal and magnetic structure of Pb(TiO)Cu$_4$(PO$_4$)$_4$.
(a) The bird's-eye view of the structure illustrating a layered crystal structure composed of two types of Cu$_4$O$_{12}$ square cupolas labeled $\alpha$ (upward) and $\beta$ (downward). The black and orange balls denote Cu and O ions, respectively.
The spin arrangement without an external magnetic field is illustrated only for the upper layer, where the red and blue arrows indicate spins with positive and negative $c$-axis ($|| Z$-axis) components, respectively. The same spin arrangement appears in all other layers.
The gray dotted line represents a unit cell. The interlayer coupling $J''$ is indicated by a black dashed line. The DM vector ($\bf D$) considered in the theoretical model is shown in the lower layer, which makes an angle $\theta$ from the $c$ axis (see text).
(b) The top view of the upper layer. The definition of the $X$, $Y$, and $Z$ axes used in the text is illustrated. Each layer is characterized by the staggered arrangement of $\alpha$ and $\beta$. When the spin arrangement is not considered (i.e., in the paramagnetic phase), $\alpha$ and $\beta$ are mutually converted by the two-fold rotational symmetry along the $X$ and $Y$ axes ($2~||~X$ and $2~||~Y$). The three dominant exchange couplings, intracluster $J_1$ and $J_2$, and intercluster $J'$ are indicated. 
\label{crystal}
}
\end{figure}

In this paper, we consider a magnetic structural unit with convex geometry known as square cupola, depicted in Fig.~\ref{design}.
It consists of four corner-sharing $MX_4$ plaquettes, where $M$ is a magnetic ion carrying a spin and $X$ is an anionic ligand. Notably, this unit can be found in a wide variety of systems ranging from minerals \cite{giester2007crystal}, salt-inclusion compounds \cite{Hwu2002}, to metal-organic hybrid systems \cite{Williams2015}.
A peculiar non-coplanar spin arrangement was recently found in the family of square cupola based antiferromagnets $A$(TiO)Cu$_4$(PO$_4$)$_4$ ($A$ = Ba, Sr, and Pb) (see, Fig.~\ref{crystal}) \cite{KKimura2016,KKimura2016b,kimura2018a}. The $ab$-plane spin component can be regarded as a magnetic quadrupole moment which is a source of linear magnetoelectric effects ---linear induction of electric polarization (magnetization) by a magnetic field (electric field). In fact, in Pb(TiO)Cu$_4$(PO$_4$)$_4$ the macroscopic electric polarization $normal$ to the convex direction (parallel to the $ab$ plane) was observed in a magnetic field applied along the $ab$ plane \cite{kimura2018a}. However, the value of the polarization is as small as $\sim 60 ~\rm{\mu C/m^2}$ at a magnetic field of 9 T, and the microscopic mechanism has not been identified. Here, we propose a new magnetoelectric response of the square cupola spin cluster; that is, an electric polarization due to the exchange striction mechanism emerges $along$ the convex direction ($c$ axis), whose sign can be controlled by a magnetic field. Targeting Pb(TiO)Cu$_4$(PO$_4$)$_4$ as an model material, we successfully verify this proposal by high-field measurements of magnetization and electric polarization, collaborated with the analysis of an effective spin model and first-principles calculations.

The rest of the paper is organized as follows.
In Sec. II, we explain our proposal for the magnetically-controllable electric polarization due to the exchange striction mechanism.
In Sec. III, we describe the crystal structure and the previously reported magnetic properties of the target material Pb(TiO)Cu$_4$(PO$_4$)$_4$.
In Sec. IV, we describe the experimental and theoretical methods used in the present study.
We present our experimental and theoretical results in Sec. V, and summarize our findings in Sec. VI.

\section{PROPOSAL FOR EXCHANGE STRICTION DRIVEN ELECTRIC POLARIZATION}
Our proposed idea is summarized in Fig.~\ref{design}.
It should be noted that, because of the convex geometry of the square cupola, there is a structurally-fixed electric dipole along the convex direction (direction of the $+z$ direction in Fig.~\ref{design}). 
This electric dipole is not controllable by an external field and therefore out of our scope.

Now we explain the onset of an electric dipole induced by the exchange striction mechanism \cite{Sergienko2006ferroelectricity}. Let us first consider a simple case, where the antiferromagnetic spin arrangement shown in Fig.~\ref{design}(a) emerges on the square cupola.
According to the Goodenough-Kanamori-Anderson rules for superexchange interactions, the metal-ligand-metal bonding angle would become closer to 180$^{\circ}$ by a shift of ligands from their original position, known as exchange striction. Then, this shift of the negatively-charged ligand gives rise to an electric dipole along its counter-direction. 
In this exchange striction mechanism, the electric dipole ${\bf p}_{ij}$ generated by the neighboring two magnetic ions $i$ and $j$ is proportional to $\langle{\bf  S}_{i} \cdot  {\bf S}_{j} \rangle$, the expectation value of the inner product of their spin operators ${\bf S}_i$ and ${\bf S}_j$. By summing ${\bf p}_{ij}$ for all the nearest-neighbor spin pairs, we obtain the electric polarization ${\bf P}^{\rm SC}$ from the square cupola unit, formulated as
\begin{eqnarray}
{\bf P}^{\rm SC} = \sum_{\langle i,j \rangle} {\bf p}_{ij} = A\sum_{\langle i,j \rangle} {\bf e}_{ij} \langle{\bf  S}_{i} \cdot  {\bf S}_{j} \rangle.
\label{dipole}
\end{eqnarray}
Here, $A~> 0$ is the constant depending on microscopic details of superexchange interactions and ${\bf e}_{ij}$ is the unit vector that determines the direction of ${\bf p}_{ij}$.
${\bf e}_{ij}$ points from the ligand site shared by magnetic ions $i$ and $j$ to the center of the bond between these magnetic ions.
Significantly, owing to the convex geometry of the square cupola, every spin pair cooperatively generates ${\bf p}_{ij}$ pointing to the convex direction with a small tilting. The tilted components normal to the convex direction cancel out with each other. As a result, a finite ${\bf P}^{\rm SC}$ appears along the convex direction.

It is immediately predicted from Eq.~(\ref{dipole}) that, when the spins align ferromagnetically [Fig.~\ref{design}(c)], the induced ${\bf P}^{\rm SC}$ is antiparallel to the convex direction. No ${\bf P}^{\rm SC}$ is expected at the intermediate state with a 90$^\circ$ spin arrangement [Fig.~\ref{design}(b)]. As a result, ${\bf P}^{\rm SC}$ can be continuously controlled from the positive to negative direction by changing the spin arrangement with an applied magnetic field. These considerations suggest that the square cupola unit is a promising structural unit carrying a magnetically-controllable polarization due to the exchange striction mechanism. It is therefore expected that a material consisting of square cupola units deserves to be explored for a large magnetoelectric coupling. 

Although we explained our proposal using the very simple example of spin arrangements shown in Fig.~\ref{design}, this proposal can be easily extended to more complex cases such as a non-coplanar spin arrangement. Moreover, not only ordered components of spins (i.e., $\langle{\bf  S}_{i} \rangle \neq 0$), but also quantum spin fluctuations can induce ${\bf P}^{\rm SC}$. An extreme example is a quantum mechanical nonmagnetic singlet state which has $\langle{\bf  S}_{i} \rangle = 0$, but $\langle{\bf  S}_{i} \cdot  {\bf S}_{j} \rangle \neq 0$. Therefore, the present idea can be applied to various kinds of spin states.

\section{TARGET MATERIAL $\bf{Pb(TiO)Cu_4(PO_4)_4}$}
In order to verify our proposal, we have targeted Pb(TiO)Cu$_4$(PO$_4$)$_4$ as a model material \cite{kimura2018a}. The crystal structure belongs to a tetragonal nonpolar space group $P42_12$, which consists of a two-dimensional staggered array of upward and downward magnetic square cupola clusters Cu$_4$O$_{12}$, as shown in Figs.~\ref{crystal}(a) and \ref{crystal}(b). The upward and downward square cupolas, which we call $\alpha$ and $\beta$, respectively, are mutually converted by symmetry operation 2 (two-fold rotation) along the [110] and [-110] axes depicted in Fig.~\ref{crystal}(b). In the following, we refer to the [110], [-110], and [001] axes as $X$, $Y$, and $Z$ axes, respectively.
The intercluster interaction ($J'$) within the array is expected to be weaker than the nearest neighbor $J_1$, so that the material can be regarded as a weakly coupled Cu$_4$O$_{12}$ system. Of the various systems with square cupolas \cite{giester2007crystal,Hwu2002,Williams2015,KKimura2016,KKimura2016b,kimura2018a}, this material benefits from an availability of sizable single crystals. Moreover, an effective spin Hamiltonian developed for isostructural Ba(TiO)Cu$_4$(PO$_4$)$_4$ \cite{Kato2017} would be applicable to the present material. 
These benefits enable a detailed comparison between experiments and theory that is crucial for microscopic understanding of magnetoelectric couplings.

Pb(TiO)Cu$_4$(PO$_4$)$_4$ undergoes a magnetic ordering at  $T_{\rm N} \approx ~ 7 $ K
without an external magnetic field. As illustrated in Figs.~\ref{crystal}(a) and \ref{crystal}(b), the four spins of each square cupola form a peculiar ``$two$-$in$, $two$-$out$'' arrangement, where the $Z$-axis components of spins align in the antiferromagnetic up-down-up-down manner, while the $XY$-plane components rotate by 90$^\circ$. 
As mentioned above, the $XY$-plane spin components can be regarded as a magnetic quadrupole moment providing a source for the linear magnetoelectric effect. The magnetic-field-induced electric polarization was indeed observed in the material, and its direction is parallel to the $XY$-plane (e.g., The electric polarization appears along the $Y$ axis when a magnetic field is applied along the $X$ axis).

When Eq.~(\ref{dipole}) is applied to the spin arrangement, one can expect that the $Z$-axis spin components induce a finite ${\bf P}^{\rm SC}$ along the $Z$ axis in each square cupola. ${\bf P}^{\rm SC}$ for $\alpha$ and $\beta$ is defined as ${\bf P}^{\alpha}$ and ${\bf P}^{\beta}$, respectively. In Pb(TiO)Cu$_4$(PO$_4$)$_4$, however, a macroscopic electric polarization ${\bf P} = {\bf P}^{\alpha} + {\bf P}^{\beta}$ was not observed because a relation ${\bf P}^{\alpha} = -{\bf P}^{\beta}$ is enforced due to the staggered arrangement of square cupolas $\alpha$ and $\beta$. Since the resultant staggered antiferroelectric polarization cannot be measured directly, it is insufficient to verify our present proposal. In the present study, we discover macroscopic $\bf P$ along the $Z$ axis in a newly found magnetic-field-induced phase, where the relation  ${\bf P}^{\alpha} = -{\bf P}^{\beta}$ is broken, as we will see below.

\begin{figure*}[htbp]
\includegraphics[width=18cm]{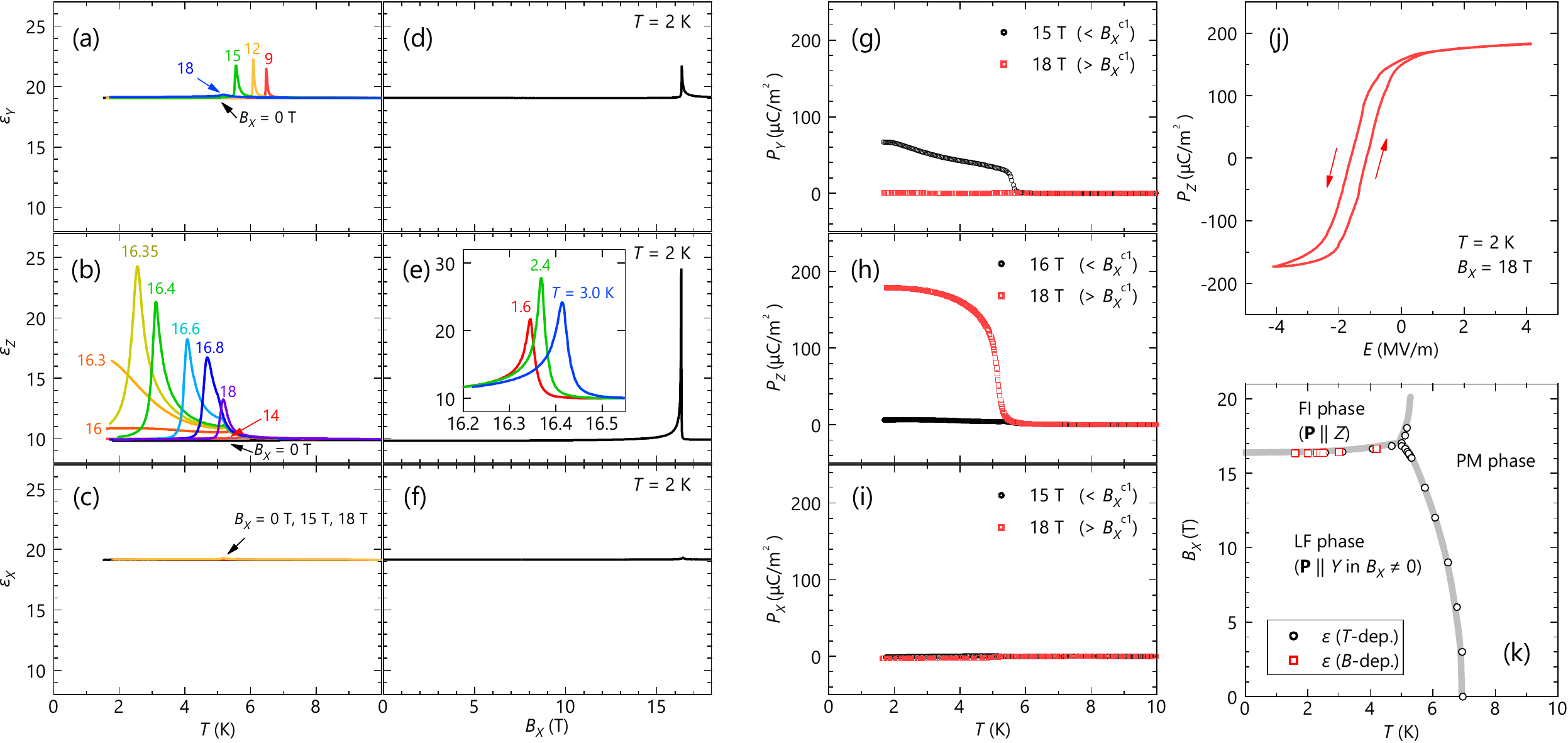}
\caption{
Dielectric constant ($\varepsilon$) and electric polarization ({\bf P}) in a magnetic field applied along the $X$ axis ($B_X$).
(a)-(c) The temperature ($T$) dependence of $\varepsilon$ along the $Y$ ($\varepsilon_Y$) (a), $Z$ ($\varepsilon_Z$) (b), and $X$ ($\varepsilon_X$) (c) axes at various strength of $B_X$.
(d)-(f) The $B_X$ dependence of $\varepsilon_Y$ (d), $\varepsilon_Z$ (e), and $\varepsilon_X$ (f) at $T$ = 2 K. The $B_X$-induced phase transition occurs at a critical field $B_X^{\rm c1} \sim 16.4$ T as indicated by the sharp peaks in (d) and (e). The inset of (e) shows $\varepsilon_Z$ at selected $T$s near $B_X^{\rm c1}$. 
(g)-(i) The $T$ dependence of {\bf P} along the $Y$ ($P_Y$) (g), $Z$ ($P_Z$) (h), and $X$ ($P_X$) (i) axes at $B_X < B_X^{\rm c1}$ (black circle) and $B_X > B_X^{\rm c1}$ (red square). The measurements were performed without an applied electric field ($E$) after the sample was cooled with $E = 0.67$ MV/m.
(j) $PE$ hysteresis loop along the $Z$ axes at $B_X=18$ T and at $T = 2$ K. The linear contribution $\varepsilon_Z E$, where $\varepsilon_Z\approx  10 $ obtained from the data in the panel (e) at $B_X=18$ T and at $T = 2$ K, is subtracted.
(k) The $B_X$ versus $T$ phase diagram determined by the anomalies seen in the $T$- and $B_X$-dependence of $\varepsilon_Z$.
\label{dielectric}
}
\end{figure*}

\section{EXPERIMENTAL AND THEORETICAL METHODS}
Single crystals of Pb(TiO)Cu$_4$(PO$_4$)$_4$ were grown by the slow cooling method \cite{kimura2018a}. Powder X-ray diffraction (XRD) measurements on crushed single crystals confirmed a single phase. The crystal orientation was determined by the Laue X-ray method.
A superconducting magnet system up to 18 T and down to 1.6 K at the Tohoku University was used for measurements of dielectric constant $\varepsilon$ and $\bf P$. For the measurements of $\varepsilon$ and $\bf P$, single crystals were cut into thin plates and subsequently an electrode was formed by painting silver pastes on a pair of the widest surfaces. Using an $LCR$ meter (Agilent E4980), $\varepsilon$ was measured at an excitation frequency of 100 kHz. $\bf P$ was obtained by integrating a pyroelectric current measured with an electrometer (Keithley 6517).
The measurements of magnetization $\bf M$ and $\bf P$ up to $\sim56$ T were performed using a multilayer pulse magnet installed at the International MegaGauss Science Laboratory of the Institute for Solid State Physics at The University of Tokyo. $\bf M$ was measured by the conventional induction method using  coaxial pickup coils. $\bf P$ was obtained by integrating the polarization current \cite{mitamura2007dielectric}. 
Multifrequency electron spin resonance (ESR) measurements (600-1400 GHz) in pulsed magnetic fields were performed to obtain the $g$ values for the field directions along the [100], [110], and [001] directions. The $g$ values were found to be isotropic within the experimental accuracy: $g\sim 2.20(5)$ for the three directions. This value is similar to that of the isostructural compound Ba(TiO)Cu$_4$(PO$_4$)$_4$ \cite{Kato2017}.
The crystal structures displayed in this paper were drawn using VESTA software \cite{Momma2011}.

To understand magnetoelectric properties obtained by the experiments, we carried out cluster mean-field (CMF) calculations of an effective spin model.
In the analysis, we consider an effective spin model associated with $S=1/2$ degree of freedom of a Cu$^{2+}$-ion which was previously constructed in Ref.~\cite{Kato2017}.
The model and the parameter setting are described in detail in Sec. V C.
Even though the effective spin model is simple, an unbiased treatment is still difficult. 
Accordingly, we analyze the model using the CMF approximation.
In the CMF treatment, the intracupola interactions are dealt with by the exact diagonalization so that the quantum effects within a cupola are fully taken into account, while the intercupola interactions are dealt with by the conventional mean-field approximation; that is, ${\bf S}_i \cdot {\bf S}_j $ is decoupled as 
${\bf S}_i \cdot {\bf S}_j \simeq \langle {\bf S}_i \rangle \cdot {\bf S}_j + {\bf S}_i \cdot \langle {\bf S}_j \rangle - \langle {\bf S}_i \rangle \cdot \langle {\bf S}_j \rangle$.
This approximation is suitable for cluster based magnetic insulators with weaker intercluster interactions.
Indeed, we successfully reproduced the magnetization curve and the dielectric anomaly observed in an
isostructural of our target material, Ba(TiO)Cu$_4$(PO$_4$)$_4$~\cite{Kato2017}.

Density functional theory (DFT) calculations were also performed to estimate the magnitude of the magnetically-induced electric polarization. The VASP (Vienna ab initio simulation package) \cite{Kresse1996} was used with a projector-augmented wave basis set. The electronic exchange and correlation were described by the Perdew-Burke-Ernzerhof generalized gradient approximation (PBE-GGA) \cite{PBE1996}. The DFT + $U$ method \cite{Liechtenstein1995} was used for the correction of strongly correlated Cu-3$d$ states, where the on-site Coulomb repulsion $U_{\rm eff}$ was set to 4 eV \cite{kimura2018a}. We first fully optimized the crystal structure of Pb(TiO)Cu$_4$(PO$_4$)$_4$ starting from the experimental structure and then optimized atomic coordinates at each spin configuration given by the model calculations under the magnetic fields. The magnetically-induced polarization was finally evaluated as the change of the polarization calculated by the Berry phase method \cite{smith1993theory,Resta1994macroscopic}.

\section{RESULTS AND DISCUSSION}
\subsection{Experiments in 18 T superconducting magnet}
Figure~\ref{dielectric} summarizes the dielectric properties in the magnetic field applied along the $X$ axis ($B_X$) below 18 T.  As seen in Figs.~\ref{dielectric}(a) and \ref{dielectric}(g), the application of $B_X$ induces a sharp peak in $\varepsilon$ along the $Y$ axis ($\varepsilon_Y$), accompanying the onset of a finite $\bf P$ along the same direction ($P_Y$). This behavior is consistent with the previous report \cite{kimura2018a}, which results from the above-mentioned linear mangetoelectric effect due to the quadrupole type spin arrangement. 
As we will see in the following subsections, the exchange striction mechanism given by Eq.$~$(\ref{dipole}) is able to reproduce the $B_X$-induced $P_Y$.

By further increasing $B_X$ above 12 T, the $\varepsilon_Y$ peak is suppressed and then completely disappears at $B_X = 18$ T, as shown in  Fig.~\ref{dielectric}(a). Correspondingly, a new anomaly appears in $\varepsilon$ along the $Z$ axis ($\varepsilon_{Z}$). These results show that a phase transition between two different magnetoelectric states is induced by the application of $B_X$. The $B_X$ dependence of $\varepsilon_Y$ and $\varepsilon_Z$ [Figs.~\ref{dielectric}(d) and (e)] reveals that the transition from the low-field (LF) to the field-induced (FI) phase occurs at the critical field $B_X^{\rm c1} = 16.4$ T at $T=2$ K. Notably, the $B_X$ dependence of  $\varepsilon_Z$ reveals a remarkably large magnetodielectric effect, defined by $[\varepsilon_{Z}(B)-\varepsilon_{Z}(0)]/\varepsilon_{Z}(0)$, with the highest value of 180 \% at 2.4 K, as shown in the inset of Fig.~\ref{dielectric}(e).
This value is comparable to the ``colossal'' magnetodielectric effect in some magnetoelectric multiferroic materials, e.g., $\sim 100$ \% for DyMn$_2$O$_5$ \cite{hur2004colossal} and $\sim 500$ \% for DyMnO$_3$ \cite{goto2004ferroelectricity}, which
indicates a very strong magnetoelectric coupling in the present material Pb(TiO)Cu$_4$(PO$_4$)$_4$. The $B_X$ versus $T$ phase diagram constructed from $\varepsilon_Z$ anomalies is drawn in Fig.~\ref{dielectric}(k). 

Measurements of $B_X$ effects on $\bf P$  have elucidated the origin of the gigantic anomalies in $\varepsilon$. Figures~\ref{dielectric}(g) and \ref{dielectric}(h) show the $T$ profiles of $\bf P$ along the $Y$ ($P_Y$) and $Z$ axes ($P_Z$), respectively, at $B_X$ below and above $B_{X}^{c1}$. These data were taken on warming without an applied electric field ($E$) after the sample was cooled with $E = 0.67$ MV/m. As shown in Fig.~\ref{dielectric}(g), by applying $B_{X}=18$ T $> B_{X}^{c1}$, $P_Y$ is strongly suppressed to almost zero. In sharp contrast, $P_{Z}$ appears [Fig.~\ref{dielectric}(h)] and its onset $T$ is fully consistent with the phase diagram [Fig.~\ref{dielectric}(k)].
In addition, no anomaly associated with the transition is seen in $\varepsilon$ and $\bf P$ along the $X$ axis ($\varepsilon_{X}$, $P_{X}$), as shown in Figs.~\ref{dielectric} (c), (f), and (i).
These results demonstrate that the direction of $\bf P$ is flopped from the $Y$ to the $Z$ axis by applying $B_X$.
Notably, the direction of $\bf P$ in the FI phase coincides with the one predicted by our proposal.
However, as discussed in Sec. III, it is not straightforward to understand an origin for the emergence of the finite $P_Z$ when considering the staggered arrangement of square cupolas $\alpha$ and $\beta$ that are related by the two-fold rotational operation about the $X$ and $Y$ axes (Fig.~\ref{crystal}). As we will see later, a finite $P_Z$ originates from breaking of this symmetry due to a spin arrangement that appears in the FI phase.

In order to examine whether the FI phase is ferroelectric, we have measured a $PE$ hysteresis curve at $T=2$ K. As shown in Fig.~\ref{dielectric}(j), $P_Z$ can be reversed by applying $E$, evidencing a ferroelectricity. Note that the hysteresis curve is highly asymmetric with respect to $E$. It is known that this behavior is observed in ferroelectric thin films, which is called an imprint effect \cite{warren1995voltage}.
Consistently, a finite $P_Z$ is observed even without the $E$-cooling procedure (not shown). The origin of the imprint effect is unclear and left for future work.

\subsection{Experiments using a pulse magnet up to 56 T}

\begin{figure}[tp]
\includegraphics[width=8.6cm]{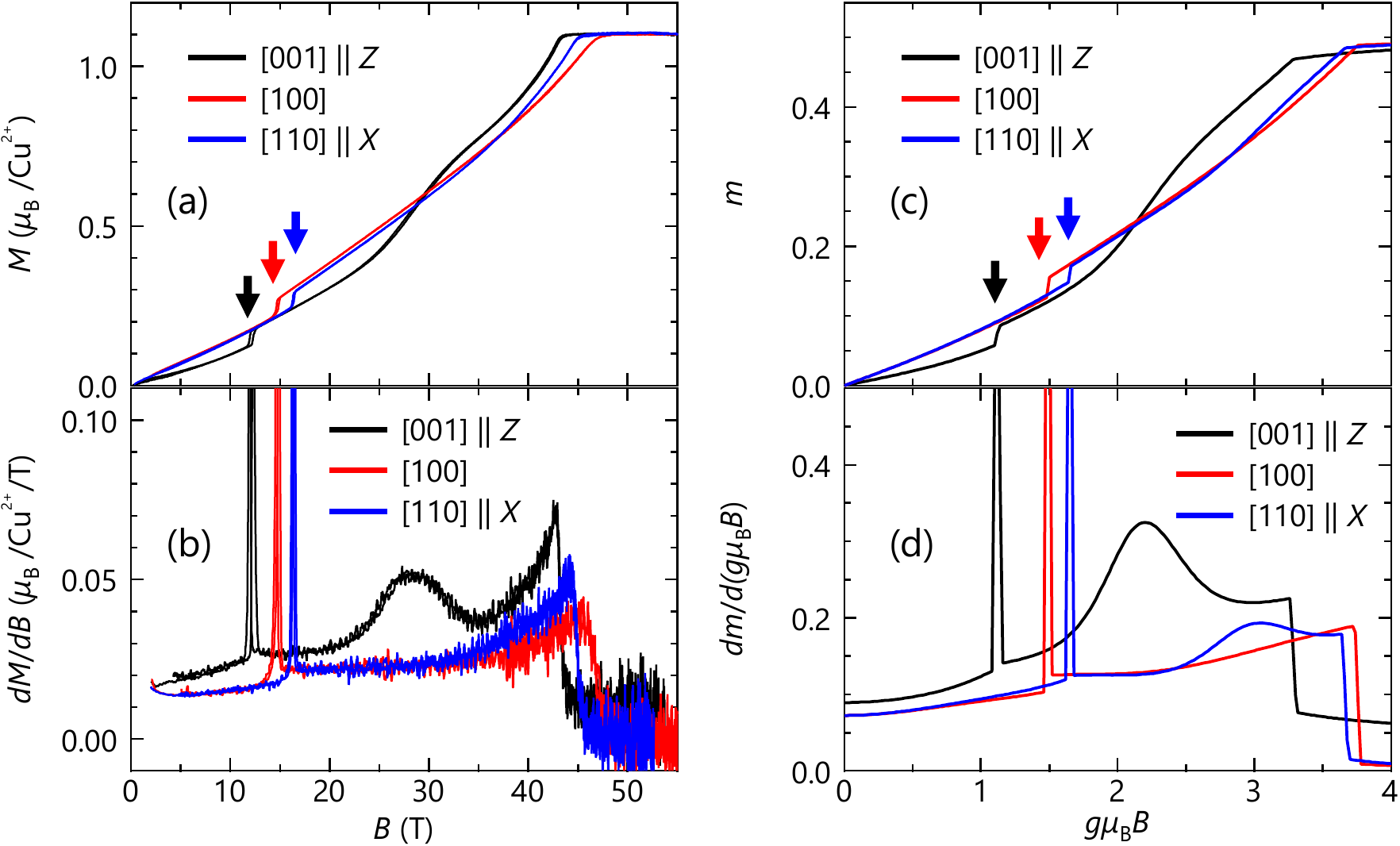}
\caption{
(a),(b) Magnetic field ($\bf B$) dependence of magnetization ($\bf M$) (a) and its field derivative $dM/dB$ (b) in the experiment at $T = 1.4$ K for $\bf B$ applied along the $Z$, [100], and $X$ axes. 
(c),(d) The magnetic field ($g\mu_{\rm B}B$) dependence of magnetization ($m$) (c) and its field derivative $dm/d(g\mu_{\rm B}B)$ (d) in the theory for the model in Eq.~(2).
$dM/dB$ in (b) and $dm/d(g\mu_{\rm B}B)$ in (d) for the $Z$ axis is shifted by 0.01 and 0.05, respectively, for clarity.
\label{mag}
}
\end{figure}

To confirm the link between the magnetism and the observed ferroelectricity, we performed high-field magnetization measurements up to 56 T using a pulse magnet. 
We show in Fig.~\ref{mag}(a) the magnetization curve in $B_X$ ($M_X$) at $T = 1.4$ K. The $M_X$ curve initially exhibits a jump at 16.4 T, which coincides well with the critical field $B_X^{\rm c1}$. This indicates that the observed flop of $\bf P$ is associated with the $B_X$-induced magnetic phase transition. Then, the $M_X$ curve gradually increases and shows a saturation above $B_X^{\rm c2} \approx 45$ T, which corresponds to a transition to a fully-polarized (FP) phase.

We have also measured magnetization curves for the field applied along the [100] ($M_{\rm [100]}$) and $Z$ ($M_Z$) axes as they provide a critical information for constructing an effective spin model, as shown later. We find that $M_{\rm [100]}$ and $M_Z$ also show an abrupt jump at $B_{[100]}^{c1}=14.8$ T and $B_Z^{c1}=12.3$ T, respectively. The saturation fields are $B_{[100]}^{c2}=47.2$ T and $B_Z^{c2}=43.4$ T. Importantly, the $M_Z$ curve shows another weak anomaly at around 28 T, which is seen as a broad hump in its field derivative [Fig.~\ref{mag}(b)]. These characteristic features provide a critical test to check the validity of the spin model.

\begin{figure}[t]
\includegraphics[width=7cm]{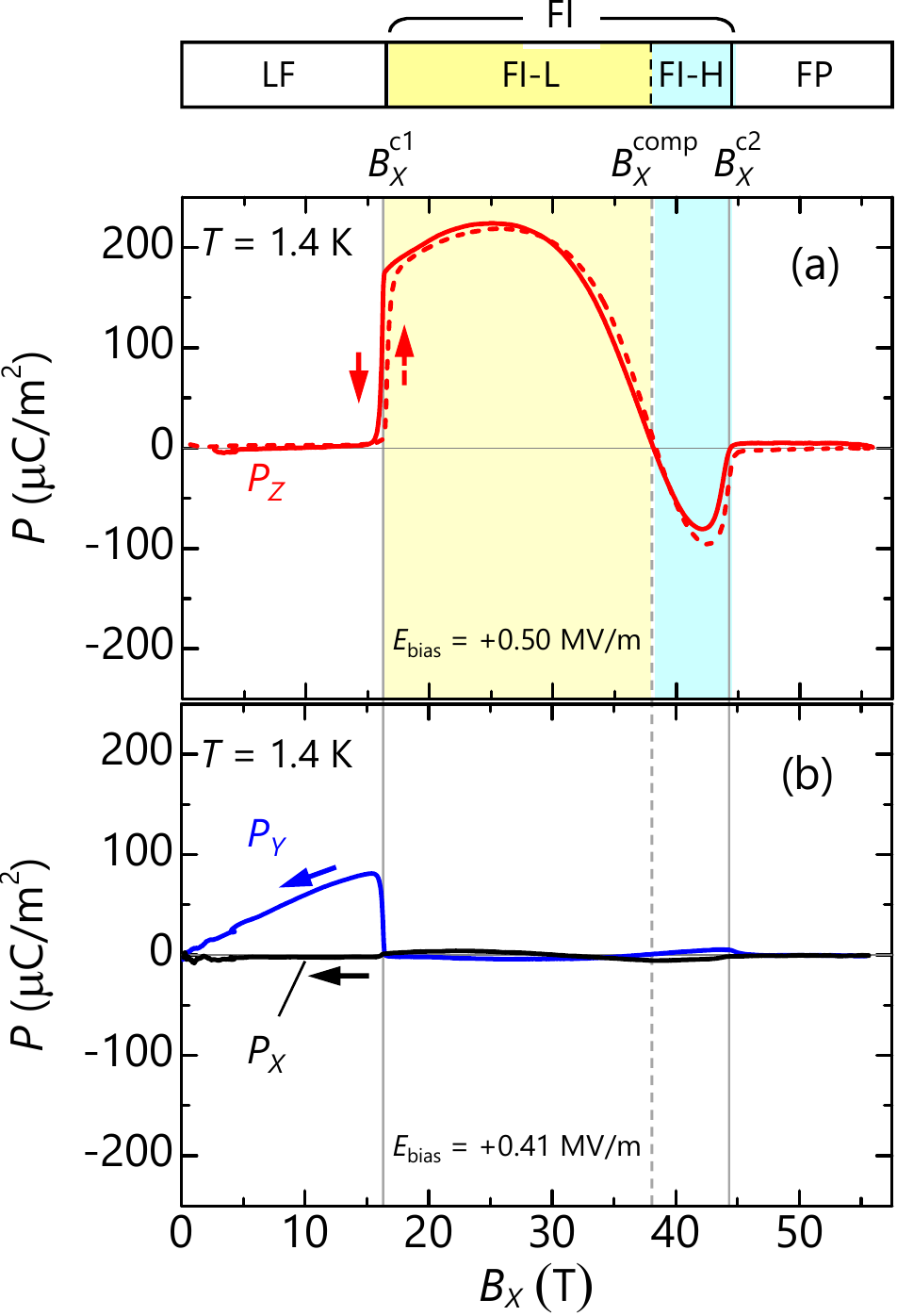}
\caption{
Electric polarization along the $X$ ($P_X$), $Y$ ($P_Y$), and $Z$ ($P_Z$) axes as a function of a magnetic field applied along the $X$ axis ($B_X$) measured using a pulse magnet. 
(a) $P_Z$ measured with an electric field $E = 0.50$ MV/m during the $B_X$-increasing (red dashed curve) and $B_X$-decreasing (red solid curve) processes. 
(b) $P_Y$ (blue solid curve) and $P_X$ (black solid curve) measured with $E = 0.41$ MV/m during the $B_X$-decreasing process.
LF, FI, and FP mean the low-field phase, field-induced phase, and fully-polarized phase, respectively. FI-L and FI-H denote the regions in the FI phase below and above the compensation field $B_X^{\rm comp}$, respectively. 
\label{pulseP}
}
\end{figure}

Now, we turn to pulse magnet measurements of $\bf P$ up to 56 T, in order to examine whether the $B_X$ dependence of $P_Z$ in the FI phase follows our proposed idea in Fig.~\ref{design}.
As shown in Fig.~\ref{pulseP}(a), after showing a broad maximum around 25 T, $P_Z$ gradually decreases and exhibits a sign reversal at around 37 T. Then, $P_Z$ vanishes in the FP phase above $B_X^{\rm c2} \approx 45$ T.
No finite component of $P_X$ and $P_Y$ is seen in the FI phase for $B_X^{\rm c1} < B_X  < B_X^{\rm c2}$.
This means that the continuous $P_Z$ reversal at around 37 T is not due to a $\bf P$ rotation, but a change of the magnitude passing through $P_Z = 0$. Qualitatively, this behavior is in agreement with our proposal in Fig.~\ref{design}. Here, we define $B_X \sim 37$ T for the $P_Z$ reversal as a compensation magnetic field ($B_X^{\rm comp}$), in analogy with a so-called compensation temperature for ferrimagnets at which a temperature-induced continuous magnetization reversal occurs. 

It should be noted that no anomaly is seen in the $M_X$ curve at $B_X \sim B_X^{\rm comp}$ [Fig.~\ref{mag}(a)]. This indicates that the observed continuous $P_Z$ reversal is associated with neither a phase transition nor domain switching, and thus accompanies no intrinsic hysteresis. Although the $B$-increasing and decreasing data [Fig.~\ref{pulseP}(a)] do not perfectly collapse on top with each other, it must be due to a fast $\bf B$ sweep in the pulse field measurements. 
In sharp contrast, a $\bf P$ reversal in most of magnetically-induced ferroelectrics is associated with either a metamagnetic transition or domain switching. The resultant large energy barrier between different magnetoelectric states causes a large hysteresis, which causes an undesirable energy loss in devices such as magnetoelectric sensors and oscillators. Therefore, a non-hysteresis feature of the $\bf P$ reversal in the present material may be useful for these applications.

\subsection{Quantum spin-1/2 model}
\subsubsection{Model and parameter setting}
The effective quantum spin-1/2 model was previously developed for the isostructural material Ba(TiO)Cu$_4$(PO$_4$)$_4$, which quite well reproduces the experimental magnetization curves \cite{Kato2017}. Therefore, it is expected that the model also has ability to explain the experimental results of the present material Pb(TiO)Cu$_4$(PO$_4$)$_4$. In this model, we take into account four dominant symmetric exchange interactions: the intracupola exchange interactions $J_1$ and $J_2$, together with the two intercupola interactions within a layer $J'$ and between neighboring layers $J''$  (Fig.~\ref{crystal}). In addition, we also take into account an antisymmetric DM interaction at $J_1$ bonds [Fig.~\ref{crystal}(a)].
The Hamiltonian can be written as
\begin{eqnarray}
\!\!\!\!\!\!\!\!\!
\mathcal{H}
=\sum_{\langle i,j \rangle}
\left[
J_1 {\bf S}_i \cdot {\bf S}_j 
-{\bf D}_{ij} \cdot 
\left({\bf S}_i \times {\bf S}_j \right)
\right]
+J_2 \!\!\sum_{\langle \langle i,j \rangle \rangle}
{\bf S}_i \cdot {\bf S}_j \nonumber \\
+J' \sum_{(i,j)} {\bf S}_i \cdot {\bf S}_j 
+J'' \!\!\sum_{[(i,j)]} {\bf S}_i \cdot {\bf S}_j 
-g \mu_{\rm B} \sum_i {\bf B} \cdot {\bf S}_i,
\label{hamiltonian}
\end{eqnarray}
where $\bf S_{\it i}$ represents $S=1/2$ spin at site $i$. The sums for $\langle i,j \rangle$, $\langle\langle i,j \rangle\rangle$, $ (i,j)$, and $[(i,j)]$ run over $J_1$, $J_2$, $J'$, and $J''$ bonds, respectively.  For the intracupola exchange coupling constants, we adopt the estimates from the first-principles calculation: $J_1 = 3.0$ meV and $J_2 = 0.43$ meV \cite{kimura2018a}. We set $J_1$ as the unit of energy, namely, $J_1 = 1$ and $J_2 = 1/7$. On the other hand, for the intercupola $J'$, we set a larger value, $J' = 3/4$, than the first-principles estimate of $J' \simeq 0.14$, because a small $J' \lesssim 0.4$ leads to a nonmagnetic singlet state in the CMF approach; we set $J'' = -1/100$ for ferromagnetic coupling between layers \cite{kimura2018a}. The last term in Eq.~(\ref{hamiltonian}) represents the Zeeman coupling where $g$ and  $\mu_{\rm B}$ are the isotropic $g$ factor and the Bohr magneton, respectively.

In the second term in Eq.~(\ref{hamiltonian}), referring to the Moriya rules \cite{moriya1960anisotropic}, we take the DM vector $\bf D_{\it i,j}$ in the plane perpendicular to the $J_1$ bond connecting $i$ and $j$ sites with the angle $\theta_{i,j}$ from the $Z$ axis [see the green arrow in Fig.~\ref{crystal}(a)].
Note that the convex geometry of the square cupola cluster induces the in-plane component of $\bf D_{\it i,j}$. The sign of $\bf D_{\it i,j}$  is reversed between the upward ($\alpha$) and downward ($\beta$) cupolas from the symmetry. We assume uniform $\theta = \theta_{i,j}$ and $D \equiv |\bf D_{\it ij}|$, and tune the values of $\theta$ and $D$ so as to reproduce the magnetization curves obtained experimentally.

We calculate the magnetic properties of this model by a standard CMF method, in which
the intracupola interactions are treated by the exact diagonalization while
the intercupola interactions are treated by the MF approximation.  The details for the calculation procedure were described in the previous report \cite{Kato2017}.
Figures~\ref{mag}(c) and \ref{mag}(d) show the $B$ profiles of magnetization per site $m$ and its field derivative 
$dm/d(g\mu_{\rm B}B)$, respectively, obtained by the CMF calculations with $\theta = 80^{\circ}$ and $D = 1.1$. The results well reproduce the experimental data in Figs.~\ref{mag}(a) and \ref{mag}(b) in the following points. (i) $m$ shows a jumplike anomaly, whose magnetic field depends on the field direction. The critical field is consistent with experimentally observed relation, namely, $B_X^{\rm c1}$ $>$  $B_{[100]}^{\rm c1}$ $>$  $B_Z^{\rm c1}$. (ii) The saturation fields satisfy the observed relationship $B_{[100]}^{\rm c2}$ $>$ $B_X^{\rm c2}$ $>$ $B_Z^{\rm c2}$. (iii) 
$dm/d(g\mu_{\rm B}B)$ exhibits a hump at an intermediate field for $B_Z$.
Thus, our effective spin model successfully explains the experimental magnetization curves, strongly supporting the validity of our model analysis.

\begin{figure*}[t]
\includegraphics[width=18cm]{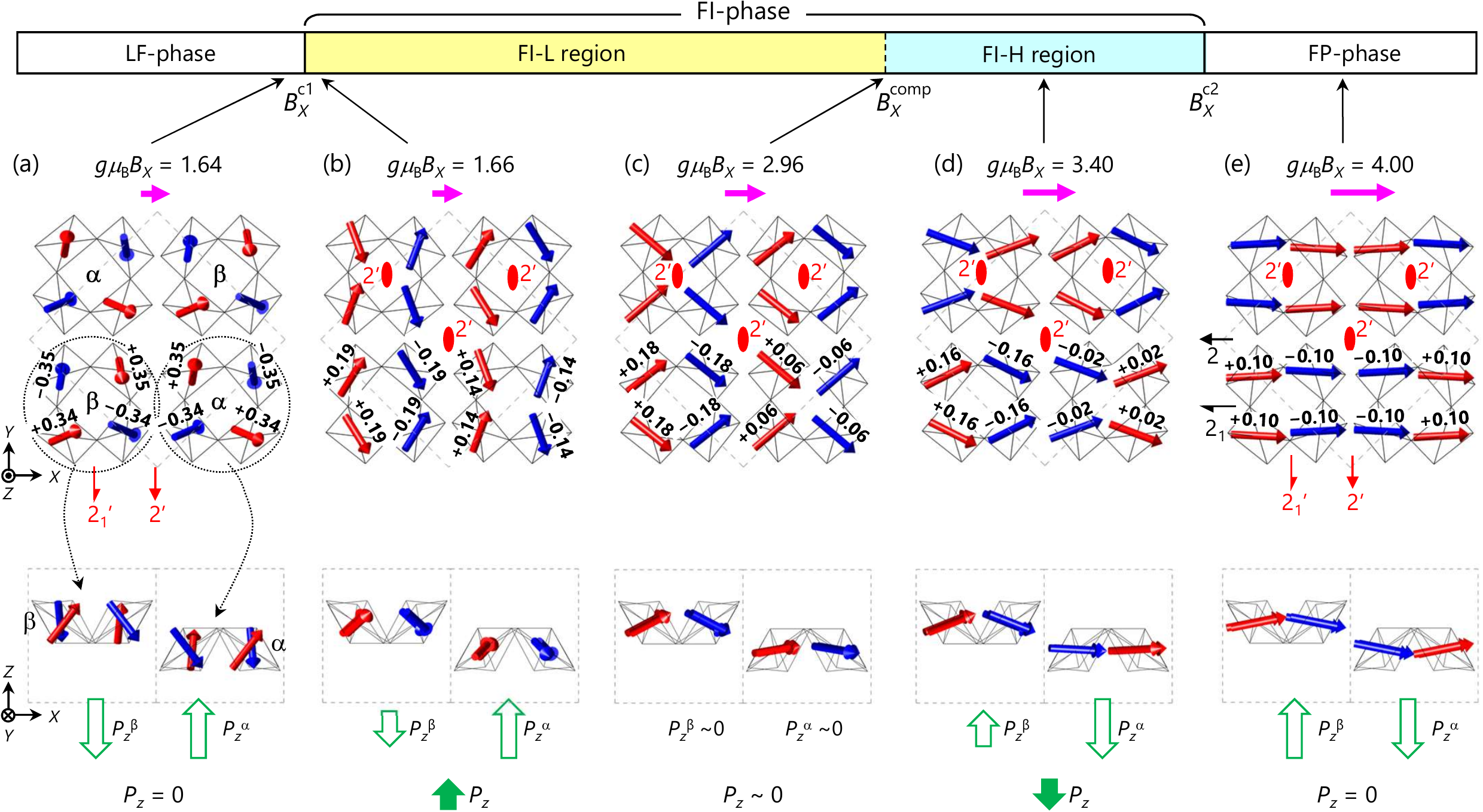}
\caption{
Calculated $B_X$-induced evolution of the ordered spin arrangement at the selected strength of $B_X$, (a) $g\mu_{\rm B}B_X = 1.64$ just below the metamagnetic transition field $B_X^{\rm c1}$, (b) $g\mu_{\rm B}B_X = 1.66$ just above $B_X^{\rm c1}$, (c) $g\mu_{\rm B}B_X = 2.96$ proximate to $B_X^{\rm comp}$, (d)  $g\mu_{\rm B}B_X = 3.40$ in-between $B_X^{\rm comp}$ and $B_X^{\rm c2}$, and (e) $g\mu_{\rm B}B_X = 4.00$ in the fully-polarized phase. The upper and lower panels are the $Z$ and $Y$-axis view of the spin arrangement, respectively.
The dashed gray line denotes the unit cell. In the $Y$-axis view, only the square cupolas marked with the dotted circles in the $Z$-axis view are indicated. The red and blue thick arrows represent spins with positive and negative $Z$-axis components, respectively. In the upper panels, values of the $Z$-axis component of each spin are given. The representative symmetry operations are also indicated. At the bottom of each panel, the electric polarization of square cupola $\alpha$ ($P_Z^{\alpha}$) and $\beta$ ($P_Z^{\beta}$) and their sum ($P_Z$), expected from the exchange striction mechanism, is schematically indicated by thick green arrows (see text for details). 
\label{spin}
}
\end{figure*}

{\subsubsection{Electric polarization}}
Before elucidating the microscopic mechanism, we discuss the onset of the net $\bf P$ in terms of symmetry of the spin arrangement calculated using the effective spin model.
In Figs.~\ref{spin}(a)-\ref{spin}(e), we summarize a $B_X$-induced evolution of the ordered spin arrangement at selected strength of $B_X$. Here, the vectors at each Cu site 
represent ($\langle{S^X_i}\rangle$, $\langle{S^Y_i}\rangle$, $\langle{S^Z_i}\rangle$), where each component $\langle{S_i^{\mu}}\rangle$, is the ordered moment along the $\mu$ axis ($\mu$ = $X$, $Y$, and $Z$).
Figure \ref{spin}(a) shows the spin arrangement in the LF phase at $B_X$ just below $B_X^{\rm c1}$. It is found that the only allowed symmetry operations are $2'$ and $2'_1$ along the $Y$ axis. The magnetic point group is therefore $2'$, which allows for the onset of $P_Y$, consistent with the experimental result [Fig.~\ref{pulseP}(b)].
Figure \ref{spin}(b) shows the spin arrangement in the FI phase just above $B_X^{\rm c1}$. It can be seen that all the spins change their orientations upon the metamagnetic transition to the FI phase.
The magnetic point group is $2'$, where $2'$ along the $Z$ axis is present at the center of each square cupola. Because of this symmetry, the polarization in each square cupola is allowed to emerge only along the $Z$ axis ($P^{\alpha}_Z$ and $P^{\beta}_Z$).
Critically, the angles made by the $XY$-plane components ($\langle{S^X_i}\rangle$ and $\langle{S^Y_i}\rangle$) of neighboring spins in square cupola $\alpha$ are different from those in $\beta$. Moreover, as denoted in the upper panel of Fig.~\ref{spin}(b), the absolute value of $\langle{S^Z_i}\rangle$ in $\alpha$ (0.14) is also different from that in $\beta$ (0.19). As a result, all the symmetry operations that mutually convert $\alpha$ and $\beta$ ($2'$ and $2'_1$ along the $Y$ axis) are broken; that is, $\alpha$ and $\beta$ are no longer symmetrically equivalent.
Therefore, the magnitude of $P^{\alpha}_Z$ and $P^{\beta}_Z$ must be different from each other, which can explain the onset of the net $P_Z$ observed in experiments [Fig.~\ref{pulseP}(a)].
By further increasing $B_X$, the spins are forced to point along the $+X$ direction, while keeping the magnetic point group $2'$  [Figs.~\ref{spin}(c) and \ref{spin}(d)].
In the FP phase [Fig.~\ref{spin}(e)], the magnetic point group is changed to $2'2'2$, where the symmetry operations that relate $\alpha$ and $\beta$ are recovered. This does not allow for $\bf P$ along any directions. Therefore, from the symmetry point of view, the calculated spin arrangement fully agrees with the onset and direction of the net $\bf P$ observed in experiments.

\begin{figure}[t]
\includegraphics[width=5.5cm]{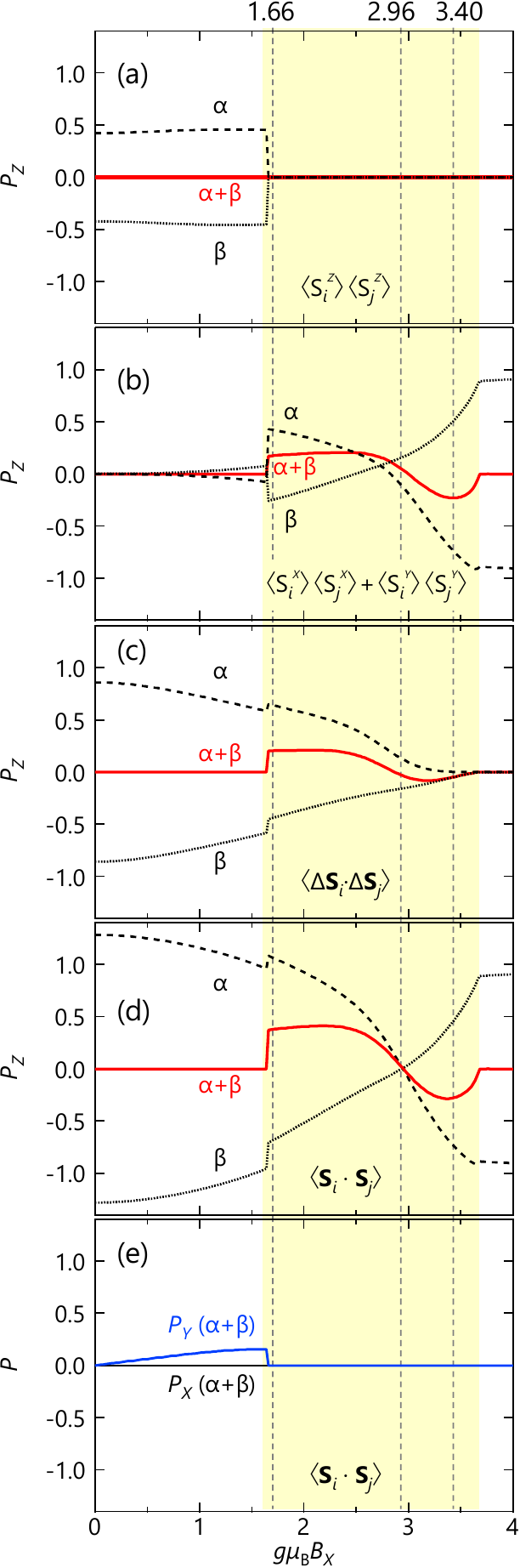}
\caption{
$B_X$ dependence of the calculated electric polarization $P_Z$ originated from
(a) $\langle{S^Z_i}\rangle \langle{S^Z_j} \rangle$,
(b) $\langle{S^{X}_i}\rangle \langle{S^{X}_j} \rangle+\langle{S^{Y}_i}\rangle \langle{S^{Y}_j} \rangle$,
(c) $\langle \Delta{\bf  S}_{i} \cdot  \Delta {\bf S}_{j} \rangle$, and
(d) $\langle {\bf  S}_{i} \cdot  {\bf S}_{j} \rangle$ terms, and
(e) $P_X$ and $P_Y$ from the $\langle {\bf  S}_{i} \cdot  {\bf S}_{j} \rangle$ term.
In (a)-(d), the contributions from square cupolas $\alpha$ and $\beta$ and their sum ($\alpha + \beta$) are shown by black dashed, black dotted, and red solid lines, respectively. The dashed vertical lines denote $B_X$ for the spin arrangements shown in Figs.~\ref{spin}(b)-\ref{spin}(d).
\label{model}
}
\end{figure}

Now, we analyze the $B_X$ dependence of $P_Z$ on the basis of the exchange striction mechanism.
Because the magnetism of Pb(TiO)Cu$_4$(PO$_4$)$_4$ is provided by quantum spins $S$ = 1/2, not only the ordered components of the spins but also quantum spin fluctuations give significant contributions to $\bf P$. We consider these two contributions separately.
To this end, we decompose $\langle{\bf  S}_{i} \cdot {\bf S}_{j} \rangle$ in Eq.~(\ref{dipole}) as
$\langle{\bf  S}_{i} \cdot {\bf S}_{j} \rangle = 
\langle{S^{X}_i}\rangle \langle{S^{X}_j} \rangle+\langle{S^{Y}_i}\rangle \langle{S^{Y}_j} \rangle+
\langle{S^Z_i}\rangle \langle{S^Z_j} \rangle +
\langle \Delta{\bf  S}_{i} \cdot  \Delta {\bf S}_{j} \rangle$.
The first three terms are ``classical'' contributions to $\bf P$, 
while the last term is a ``quantum'' contribution, where $\Delta{\bf  S}_{i} \equiv {\bf  S}_{i} - \langle{\bf  S}_{i}\rangle$ describes quantum spin fluctuations. Considering electric dipoles induced by $J_1$ bonds, we can evaluate electric polarization for each square cupola using the following formula,
\begin{eqnarray}
\!\!\!\!\!\!\!\!\!
{\bf P}^k =\sum_{\langle i,j \rangle ^{k}} {\bf e}^{k}_{ij} 
[\langle{S^{X}_i}\rangle \langle{S^{X}_j} \rangle+\langle{S^{Y}_i}\rangle \langle{S^{Y}_j} \rangle  \nonumber \\
+\langle{S^Z_i}\rangle \langle{S^Z_j} \rangle
+\langle \Delta{\bf  S}_{i} \cdot  \Delta {\bf S}_{j} \rangle],
\label{dipole2}
\end{eqnarray}
where $k = \alpha$ and $\beta$. The sum of these two gives the total electric polarization in the unit cell, ${\bf P} = {\bf P}^{\alpha} + {\bf P}^{\beta}$.
(We ignored the coefficient $A$ in Eq.~\eqref{dipole} for simplicity.)
The calculated results of the $B_X$ dependence of ${\bf P}^{\alpha}$,  ${\bf P}^{\beta}$, and ${\bf P}$ are summarized in Figs.~\ref{model}(a)-\ref{model}(e), which will be described below.

Let us first consider the classical contributions to $P_Z = P_Z^{\alpha} + P_Z^{\beta} $ in the FI-phase.
As shown in Figs.~\ref{spin}(b)-\ref{spin}(d), the $Z$-axis components $\langle{S^Z}\rangle$ of four spins within each square cupola align in the up-up-down-down manner, and their absolute values are identical to each other in the entire $B_X$ range.
According to Eq.~(\ref{dipole2}), 
the sum of $\langle{S_i^Z}\rangle \langle{S_j^Z}\rangle$ is always zero and does not induce a finite $P_Z$, as shown in Fig.~\ref{model}(a).
Therefore, at the classical level, only the $XY$-plane spin components contribute to $P_Z$. 
Notably, the $B_X$-induced evolution of the $XY$-plane components in each square cupola [upper panels of Figs.~\ref{spin}(b)-\ref{spin}(d)] is quite similar to the one in our proposal (Fig.~\ref{design}).
This leads to a continuous sign reversal of $P_Z^{\alpha}$ ($P_Z^{\beta}$) from positive to negative (negative to positive), as shown in Fig.~\ref{model}(b).
Importantly, the above-mentioned difference between the magnitude of $P_Z^{\alpha}$ and $P_Z^{\beta}$ gives rise to a net $P_Z$. Furthermore, in agreement with the experimental results, the net $P_Z$ exhibits a continuous sign reversal [red solid line in Fig.~\ref{model}(b)] due to the following reasons.
In the lower field region of the FI-phase (FI-L region) [Fig.~\ref{spin}(b)], the arrangement of the $XY$-plane components is closer to antiparallel (more antiferromagnetic) in $\alpha$ than in $\beta$. The magnitude of $P_Z^{\alpha}$ is therefore larger than that of $P_Z^{\beta}$, which gives rise to a net $P_Z$ along the $+Z$ direction [see the bottom of the panel in Fig.~\ref{spin}(b)].
On the other hand, in the higher field region of the FI-phase (FI-H region) [Fig.~\ref{spin}(d)], the  arrangement of $XY$-plane components is closer to parallel (more ferromagnetic) in $\alpha$ than in $\beta$. As a result, a net $P_Z$  appears along the $-Z$ direction, opposite to $P_Z$ in the FI-L region.
In the intermediate field region of $g\mu_{\rm B}B_X = 2.96$ [Fig.~\ref{spin}(c)], the $XY$-plane components of the neighboring spins are roughly perpendicular with each other, resulting in $P_Z \sim 0$.
The continuous sign reversal of $P_Z$ thus occurs. These results clearly demonstrate that the classical contributions of the exchange striction mechanism can explain the onset of $P_Z$ and its $B_X$-induced continuous sign reversal.

To investigate the effects of quantum spin fluctuations, we plot in Fig.~\ref{model}(c) the quantum contribution ($\langle \Delta{\bf  S}_{i} \cdot  \Delta {\bf S}_{j} \rangle$ term) to $P_Z^\alpha$, $P_Z^\beta$, and $P_Z$. 
In the FI-L region, the quantum contribution to $P_Z$ is found to be as large as the classical contribution, which demonstrates that quantum fluctuations largely enhance the ferroelectricity in Pb(TiO)Cu$_4$(PO$_4$)$_4$. 
On the other hand, in the FI-H region the quantum contribution becomes less significant, most likely due to the suppression of the quantum fluctuations by the applied strong magnetic field.

Figure \ref{model}(d) shows the total contributions to $P_Z$ of the exchange striction mechanism. The calculation result quite well reproduces the experimental $B_X$ profile of $P_Z$ in Fig.~\ref{pulseP}(a). As shown in Fig.~\ref{model}(e), the calculation result also well reproduces $B_X$ profiles of $P_X$ and $P_Y$ in Fig.~\ref{pulseP}(b). 
The $B_X$-induced $P_Y$ in the LF phase can be understood as follows.
In the case of the zero-field spin arrangement (Fig.~\ref{crystal}), the $ab$-plane component of $\bf{p}_{\it ij}$ from all the bonds cancels out completely. On the other hand, when the spin arrangement is deformed by the applied $B_X$ [Fig.~\ref{spin}(a)], the cancellation becomes incomplete, which results in the net  $P_Y$.
Therefore, the present model calculations demonstrate that the exchange striction mechanism provides an excellent description of the ferroelectricity in Pb(TiO)Cu$_4$(PO$_4$)$_4$.

We note that the maximum value of $P_Z \sim 220~\rm{\mu C/m^2}$ observed in experiments is rather small and comparable to the observed value of typical magnetically-induced ferroelectric polarization via spin-orbit couplings in inorganic oxides (order of $1-100~\mu \rm C /m^2$ \cite{kimura2012magnetoelectric}). However, this small value is obviously due to the cancellation of the sublattice polarization $P_Z^{\alpha}$ and  $P_Z^{\beta}$. If the square cupolas were arranged in a uniform manner, rather than the staggered manner as in Pb(TiO)Cu$_4$(PO$_4$)$_4$, the combined electric polarization [= $P_Z^{\alpha}~-~  P_Z^{\beta}$ in Fig.~\ref{model}(d)] would amount to $\sim 1000~\rm{\mu C/m^2}$, which is comparable to the typical polarization induced by the exchange striction mechanism (e.g., Refs.~\cite{tokunaga2008magnetic,ishiwata2011high}).
This estimate, together with the gigantic magnetodielectric effect [Fig.~\ref{dielectric}(e)], shows that the square cupola units can be considered as a building block with strong magnetoelectric couplings.

\
\subsection{Density functional calculations}
To unambiguously establish that the observed ferroelectric polarization dominantly arises from the non-relativistic exchange striction mechanism, we examine the effects of relativistic spin-orbit coupling (SOC) on the electric polarization.
To this end, we have performed DFT calculations of $P_Z$ including SOC self-consistently, whereas we have also performed extra calculations excluding SOC. The former and the latter correspond to a total $P_Z$ and $P_Z$ due to only the exchange striction mechanism, respectively. The difference between the two corresponds to the effects of SOC on $P_Z$. In the calculations, the spin arrangement obtained by the CMF analysis of the model Eq.~(\ref{hamiltonian}) was used.

\begin{figure}[t]
\includegraphics[width=8.6cm]{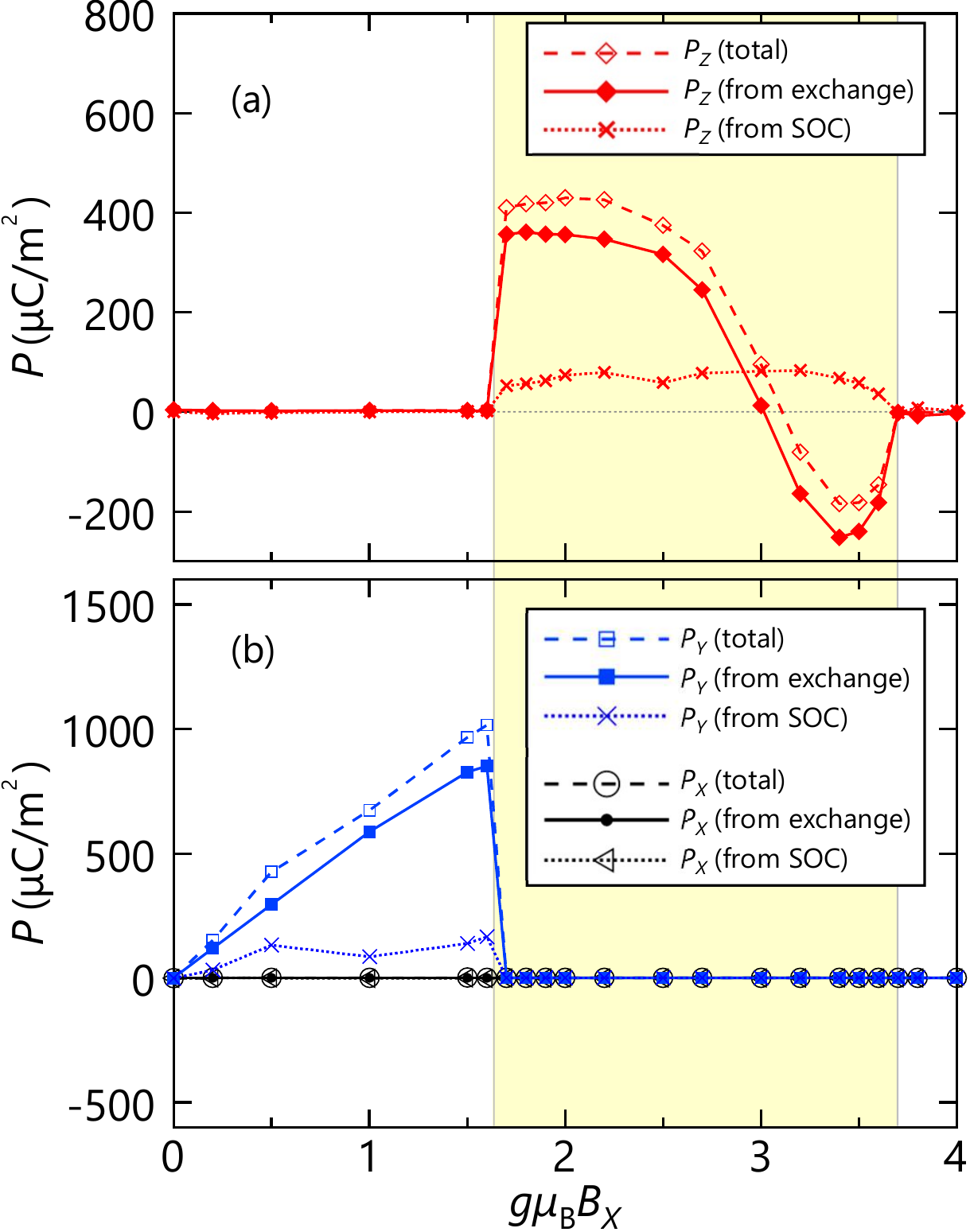}
\caption{
Density functional calculations of the electric polarization as a function of the magnetic field applied along the $X$ axis ($B_X$).
(a) The electric polarization along the $Z$ axis ($P_Z$).
To extract the effect of the spin-orbit coupling (SOC), calculations including and excluding SOC were performed.
The former and the latter correspond to the total $P_Z$ (red open diamonds) and $P_Z$ from the exchange striction (red filled diamonds), respectively. 
The red crosses denote the difference between the two, which corresponds to $P_Z$ arising from any mechanism involving SOC.
(b) The corresponding electric polarization along the $X$ axis ($P_X$) and the $Y$ axis ($P_Y$). The total $P_Y$,  $P_Y$ from the exchange striction, and $P_Y$ from SOC are denoted by blue open squares, blue filled squares, and blue crosses, respectively.
The total $P_X$, $P_X$ from the exchange striction, and $P_X$ from SOC are shown by black open circles, black filled circles, and black triangles, respectively, which are zero in the entire $B_X$ range.
\label{dft}
}
\end{figure}

Figure \ref{dft}(a) shows the results of the $B_X$ dependence of the total $P_Z$ (red open diamonds), $P_Z$ due to the exchange striction mechanism (red filled diamonds), and $P_Z$ from the effects of SOC (red crosses).
It is found that the total $P_Z$ well reproduces the $B_X$ dependence of $P_Z$ observed in experiments including the sign reversal. Importantly, $P_Z$ due to the exchange striction mechanism dominates the overall $B_X$ dependence, 
while the effects of SOC give only a minor contribution to $P_Z$ and cannot reproduce the sign reversal.
This means that any mechanisms associated with SOC, such as the spin current mechanism \cite{Katsura2005,Sergienko2006} and the metal-ligand $d$-$p$ hybridization mechanism \cite{Arima2007}, do not give an important contribution. Therefore, our DFT result clearly demonstrates that the exchange striction mechanism is dominant for the observed magnetically-induced ferroelectricity in Pb(TiO)Cu$_4$(PO$_4$)$_4$.

The maximum value of the total $P_Z$ amounts to $\sim 400~\rm{\mu C/m^2}$.
This value is in fairly good agreement with the maximum value of $P_Z \sim 220~\rm{\mu C/m^2}$ observed in experiments [Fig.~\ref{pulseP}(a)].
On the other hand, the calculated value is two orders of magnitude smaller than the exchange-striction-driven polarization calculated for other systems (e.g., $\sim 60000~\rm{\mu C/m^2}$ in HoMnO$_{3}$ \cite{picozzi2007dual}). As already mentioned, this is attributable to the cancellation of the sublattice polarization $P_Z^{\alpha}$ and  $P_Z^{\beta}$.

We also calculated the $B_X$ dependence of $P_X$ and $P_Y$, the result of which is shown in Fig.~\ref{dft}(b). It captures the qualitative features observed in experiments; that is, the onset of $P_Y$ in the LF phase and the absence of $P_X$ in the entire $B_X$ range.
However, in contrast to the quantitative agreement between
the calculated value of the total $P_Z$ and the experimental value of $P_Z$
in the FI-phase, the calculated value of the total $P_Y$ ($\sim  1000~\mu \rm C /m^2$) in the LF phase is found to be much larger than the experimental value ($<  100~ \mu \rm C /m^2$, see Fig.~\ref {pulseP}). 
This discrepancy may be explained by the difference between the spin arrangement in the calculations and experiments: The spin arrangement in the LF phase calculated by Eq.~\eqref{hamiltonian} [Fig.~\ref{spin}(a)]
is more collinear along the $Z$ axis than that proposed by the previous neutron diffraction experiments in Fig.~\ref{crystal}(a) \cite{kimura2018a}, which gives a larger 
$\langle {\bf S}_{\it i} \rangle  \cdot \langle {\bf S}_{\it j} \rangle$ and thus larger calculated $P_Y$.

\
\subsection{Electric field control of domains}
Using Pb(TiO)Cu$_4$(PO$_4$)$_4$ as a model material, we have successfully demonstrated that the magnetic square cupola units exhibit the exchange-striction-driven electric polarization, whose sign can be continuously reversed by sweeping a magnetic field. So far, the $\bf B$-induced continuous $\bf P$ reversal, not originating from a phase transition or domain switching, was observed only in a very limited number of materials including $AE$$_2$$TM$Ge$_2$O$_7$ ($AE$ = Ba, Sr, $TM$ = Co, Mn) \cite{murakawa2012comprehensive, akaki2012multiferroic} and $R$Mn$_2$O$_5$ ($R$ = Tb or Bi) \cite{hur2004electric, kim2009observation}. Little is known about unique ferroelectric properties associated with the $\bf B$-induced continuous $\bf P$ reversal. Therefore, we move to investigate ferroelectric properties of Pb(TiO)Cu$_4$(PO$_4$)$_4$ in more details, in particular a response of ferroelectric domains to an external electric field in the FI phase. Technically, measurements of $PE$ hysteresis loops at a constant $B_X$ using a pulse magnet is rather difficult. We have instead carried out a polarization measurement on sweeping $B_X$ with a various bias electric field $E_{\rm bias}$ in the range of $-2.5$ to $+3.0$ MV/m. 

\begin{figure}[t]
\includegraphics[width=8.6cm]{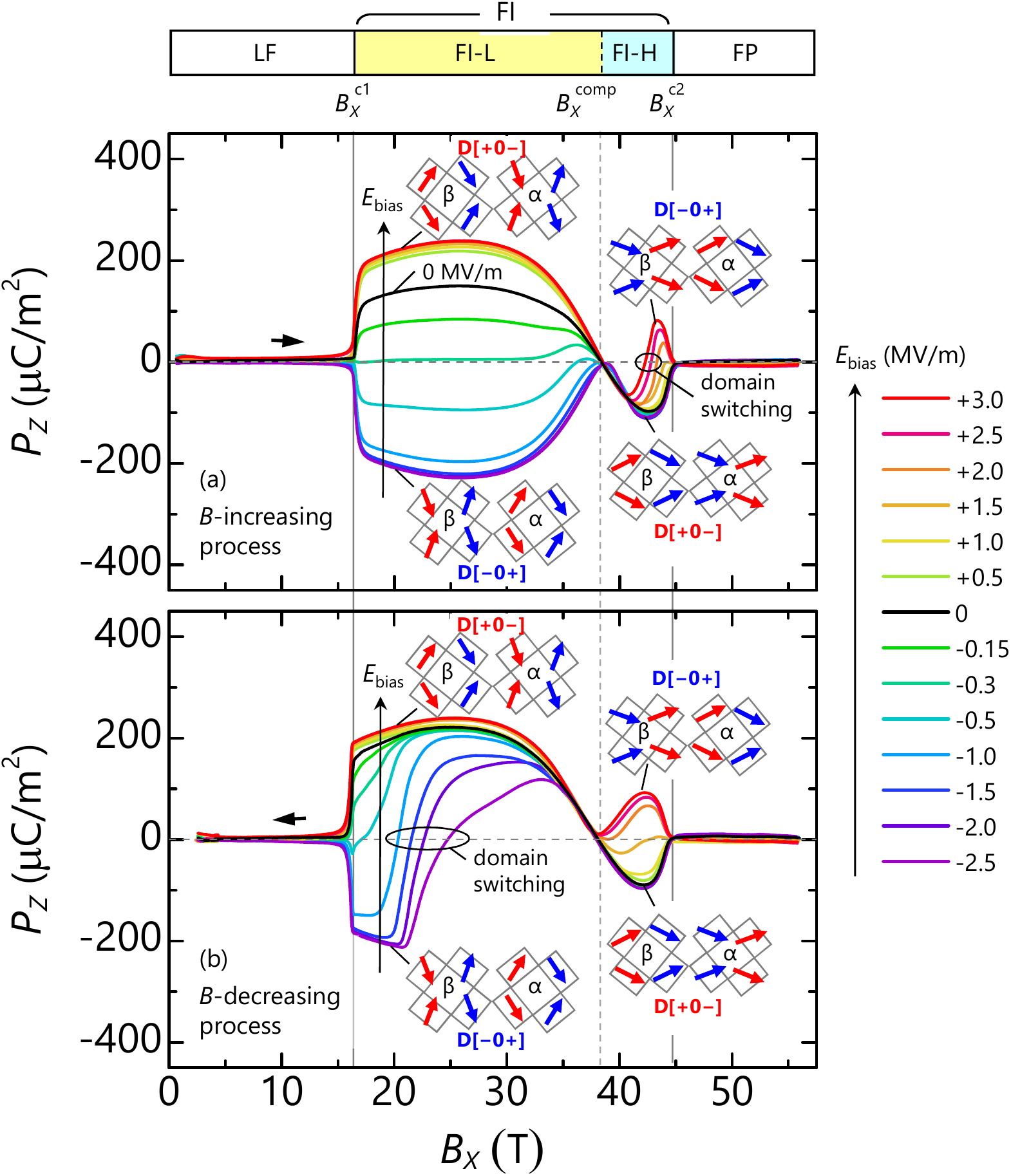}
\caption{
The $B_X$ dependence of $P_Z$ measured with various bias electric fields ($E_{\rm bias}$) during a $B_X$-increasing (a) and a subsequent $B_X$-decreasing (b) processes.
In the FI phase there are two types of domains labeled as D$[+0-]$ and D$[-0+]$; the former (latter) shows $+P_Z$ and $-P_Z$ ($-P_Z$ and $+P_Z$) in the FI-L and FI-H regions, respectively. The spin arrangements of square cupolas $\alpha$ and $\beta$ of these domains in the FI-L and FI-H regions are schematically illustrated. The red and blue arrows represent spins with positive and negative $Z$-axis components, respectively. The measurements were performed at the base temperature of the measurement system ranging from 1.4 K to 1.5 K.
\label{Edep}
}
\end{figure}

Before presenting experimental results, let us describe the ferroelectric domains in Pb(TiO)Cu$_4$(PO$_4$)$_4$. In our model calculations, two types of energetically equivalent spin arrangements are obtained in the FI-phase. These spin arrangements are mutually converted by the two-fold rotation operation about the $X$ axis, with the rotation axis passing through the center of the unit cell [see Figs.~\ref{spin}(b)-\ref{spin}(d)]. They correspond to ferroelectric domains exhibiting the opposite sign of the $B_X$ dependence of $P_Z$ to each other. Here, we define the domain which shows the positive to negative $B_X$ dependence of $P_Z$, corresponding to Fig.~\ref{model}(d) as D$[+0-]$, while another domain with the negative to positive $B_X$ dependence as D$[-0+]$. 
The spin arrangements of these domains in the FI-L and FI-H regions are schematically illustrated in Figs.~\ref{Edep}(a) and \ref{Edep}(b).

The results of $P_Z$ measured during the $B_X$ increasing and subsequent decreasing processes are separately shown in Figs.~\ref{Edep}(a) and \ref{Edep}(b), respectively. They reveal a very complicated behavior. To understand this, we first focus on the $E_{\rm bias} = 0$ MV/m data (black line). In both of the $B$-increasing and $B$-decreasing processes, positive and negative $P_Z$ appear in the FI-L and FI-H regions, respectively. This means that the above-mentioned imprint effect [see Fig.~\ref{dielectric}(j)] always stabilizes the ferroelectric domain D$[+0-]$ over the entire region of the FI phase. Therefore, the imprint effect does not directly couple to the sign of $P_Z$, but to the domain state. In this sense, D$[+0-]$ can be regarded as the imprint-stabilized domain, while D$[-0+]$ the imprint-destabilized domain.

The imprint-destabilized D$[-0+]$ state can be stabilized by the application of a strong enough $E_{\rm bias}$ across the transition from the paraelectric (LF or FP) to ferroelectric FI phases. For example, in the FI-L region during the $B$-increasing process [Fig.~\ref{Edep}(a)], the negative $P_Z$ is induced by a negative $E_{\rm bias}$. It is saturated at $E_{\rm bias} < -1.0$ MV/m, indicating the emergence of a single D$[-0+]$ state. Likewise, in the FI-H region during the $B$-decreasing process [Fig.~\ref{Edep}(b)], the positive $P_Z$ is induced by positive $E_{\rm bias}$ and becomes a nearly saturated at $E_{\rm bias} > +2.5$ MV/m, indicating a nearly single D$[-0+]$ state. Therefore, the ferroelectric domains in Pb(TiO)Cu$_4$(PO$_4$)$_4$ are highly responsive to an external electric field, which allows for the unique control of the ferroelectric polarization by the combination of magnetic and electric fields.

Finally, we consider the domain switching behavior within the ferroelectric FI phase. By the application of the positive $E_{\rm bias}$ in the $B$-increasing process [Fig.~\ref{Edep}(a)], a single domain state of D$[+0-]$ must be present at $B_X = B_X^{\rm comp}$. As $B_X$ further increases into the FI-H region, the  D$[+0-]$ domain with $-P_Z$ is switched to the D$[-0+]$ domain with $+P_Z$ by the positive $E_{\rm bias} > +0.5$ MV/m. However, this domain switching occurs only at $B_X$ far above $B_X^{\rm comp}$. In other word, a coercive electric field (i.e., $E_{\rm bias}$ for the domain switching) increases as $B_X$ approach to $B_X^{\rm comp}$. Then, nearby $B_X^{\rm comp}$ the domain switching finally becomes impossible by $E_{\rm bias}$ applied in the present study ($E_{\rm bias} \leq 3$ MV/m). A similar behavior is observed for the domain switching in the FI-L region during the $B$-decreasing process with a negative $E_{\rm bias}$ [Fig.~\ref{Edep}(b)]. As in the case of the $B$-increasing process with the positive $E_{\rm bias}$, the single domain state of D$[+0-]$ must be present at $B_X = B_X^{\rm comp}$. As $B_X$ further decreases into the FI-L region, the D$[+0-]$ domain is switched to D$[-0+]$ by a negative $E_{\rm bias} \leq -1.0$ MV/m, but it occurs only at $B_X$ far below $B_X^{\rm comp}$. Combined these results, it is expected that the coercive electric field becomes maximum at $B_X^{\rm comp}$. This is consistent with the absence of a driving force for the domain switching ($P_{Z} \times E_{\rm bias} = 0$) at  $B_X^{\rm comp}$ since $P_Z = 0$. This feature of the coercive electric field resembles well to the temperature dependence of the coercive magnetic field in ferrimagnets showing a temperature-induced magnetization reversal. In memory devices composed of ferrimagnetic materials, this compensation behavior is quite useful for robust data storage and low-field writing. Therefore, the  $\bf B$-induced compensation of the electrically-switchable $\bf P$  discovered in the present study may open a possibility of unique magnetoelectric devices. Clearly, it is essential to make a more detailed characterization of the $B_X$-dependence of the coercive electric field, such as by means of $PE$ hysteresis loop measurements with a flat pulse magnetic field.

\section{CONCLUSION}
In conclusion, we propose a convex-shaped square cupola  spin cluster as a promising structural unit hosting a large mangetoelectric coupling due to the exchange striction mechanism. Targeting Pb(TiO)Cu$_4$(PO$_4$)$_4$ as a model material, which consists of a staggered array of Cu$_4$O$_{12}$ square cupolas, our joint experimental and theoretical studies successfully verify this idea by observing the gigantic magnetodielectric effect ($\sim 180 \%$) and ferroelectricity that originates from the staggered arrangement of large electric polarization with the different magnitude.  We also discover a $B$-induced continuous reversal of the ferroelectric polarization,  which enables an unusual control of the domains by combination of electric and magnetic fields. We here emphasize that our proposal in the present study does not require a complicated, delicate balance among frustrated magnetic interactions, which are usually required in most of magnetically-induced ferroelectric materials. Owing to this simplicity, our proposed idea can be extended to other types of spin clusters with convex geometry. The present result therefore demonstrates that materials with convex-shaped magnetic structural units deserve to be explored and synthesized to achieve strong magnetoelectric couplings, which would open the door for future magnetoelectric device applications. 
\\\
\\\

\begin{acknowledgments}
We wish to thank P. Babkevich and H. M. R\o{}nnow for helpful discussions.
This work was partially supported by JSPS KAKENHI Grant Numbers JP16K05413, JP16K05449, JP17H02916, JP17H02917, JP17H01143, and JP24244058 and by the MEXT Leading Initiative for Excellent Young Researchers (LEADER).
K.Y., M. Toyoda, and T.K. were partially supported by the Center for Spintronics Research Network, Osaka University.
K.K., M.A., M.H., S.K and T.K. are partially supported by JSPS Core-to-Core Program, A. Advanced Research Networks.
Numerical calculations in this work has been done using the facilities of the Supercomputer Center, the Institute for Solid State Physics, the University of Tokyo.
High-field ESR measurements were carried out at the Center for Advanced High Magnetic Field Science in Osaka University under the Visiting Researcher's Program of the Institute for Solid State Physics, the University of Tokyo.
This work was partly performed at the High Field Laboratory for Superconducting Materials, Institute for Materials Research, Tohoku University (Project No 18H0014)
\end{acknowledgments}



\begin{thebibliography}{47}%
\makeatletter
\providecommand \@ifxundefined [1]{%
 \@ifx{#1\undefined}
}%
\providecommand \@ifnum [1]{%
 \ifnum #1\expandafter \@firstoftwo
 \else \expandafter \@secondoftwo
 \fi
}%
\providecommand \@ifx [1]{%
 \ifx #1\expandafter \@firstoftwo
 \else \expandafter \@secondoftwo
 \fi
}%
\providecommand \natexlab [1]{#1}%
\providecommand \enquote  [1]{``#1''}%
\providecommand \bibnamefont  [1]{#1}%
\providecommand \bibfnamefont [1]{#1}%
\providecommand \citenamefont [1]{#1}%
\providecommand \href@noop [0]{\@secondoftwo}%
\providecommand \href [0]{\begingroup \@sanitize@url \@href}%
\providecommand \@href[1]{\@@startlink{#1}\@@href}%
\providecommand \@@href[1]{\endgroup#1\@@endlink}%
\providecommand \@sanitize@url [0]{\catcode `\\12\catcode `\$12\catcode
  `\&12\catcode `\#12\catcode `\^12\catcode `\_12\catcode `\%12\relax}%
\providecommand \@@startlink[1]{}%
\providecommand \@@endlink[0]{}%
\providecommand \url  [0]{\begingroup\@sanitize@url \@url }%
\providecommand \@url [1]{\endgroup\@href {#1}{\urlprefix }}%
\providecommand \urlprefix  [0]{URL }%
\providecommand \Eprint [0]{\href }%
\providecommand \doibase [0]{http://dx.doi.org/}%
\providecommand \selectlanguage [0]{\@gobble}%
\providecommand \bibinfo  [0]{\@secondoftwo}%
\providecommand \bibfield  [0]{\@secondoftwo}%
\providecommand \translation [1]{[#1]}%
\providecommand \BibitemOpen [0]{}%
\providecommand \bibitemStop [0]{}%
\providecommand \bibitemNoStop [0]{.\EOS\space}%
\providecommand \EOS [0]{\spacefactor3000\relax}%
\providecommand \BibitemShut  [1]{\csname bibitem#1\endcsname}%
\let\auto@bib@innerbib\@empty
\bibitem [{\citenamefont {Astrov}(1960)}]{Astrov1960}%
  \BibitemOpen
  \bibfield  {author} {\bibinfo {author} {\bibfnamefont {D.~N.}\ \bibnamefont
  {Astrov}},\ }\href@noop {} {\bibfield  {journal} {\bibinfo  {journal} {Sov.
  Phys. JETP}\ }\textbf {\bibinfo {volume} {11}},\ \bibinfo {pages} {708}
  (\bibinfo {year} {1960})}\BibitemShut {NoStop}%
\bibitem [{\citenamefont {Dubovik}\ and\ \citenamefont
  {Tugushev}(1990)}]{Dubovik1990}%
  \BibitemOpen
  \bibfield  {author} {\bibinfo {author} {\bibfnamefont {V.~M.}\ \bibnamefont
  {Dubovik}}\ and\ \bibinfo {author} {\bibfnamefont {V.~V.}\ \bibnamefont
  {Tugushev}},\ }\href@noop {} {\bibfield  {journal} {\bibinfo  {journal}
  {Phys. Rep.}\ }\textbf {\bibinfo {volume} {187}},\ \bibinfo {pages} {145}
  (\bibinfo {year} {1990})}\BibitemShut {NoStop}%
\bibitem [{\citenamefont {Kimura}\ \emph {et~al.}(2003)\citenamefont {Kimura},
  \citenamefont {Goto}, \citenamefont {Shintani}, \citenamefont {Ishizaka},
  \citenamefont {Arima},\ and\ \citenamefont {Tokura}}]{Kimura2003}%
  \BibitemOpen
  \bibfield  {author} {\bibinfo {author} {\bibfnamefont {T.}~\bibnamefont
  {Kimura}}, \bibinfo {author} {\bibfnamefont {T.}~\bibnamefont {Goto}},
  \bibinfo {author} {\bibfnamefont {H.}~\bibnamefont {Shintani}}, \bibinfo
  {author} {\bibfnamefont {K.}~\bibnamefont {Ishizaka}}, \bibinfo {author}
  {\bibfnamefont {T.}~\bibnamefont {Arima}}, \ and\ \bibinfo {author}
  {\bibfnamefont {Y.}~\bibnamefont {Tokura}},\ }\href@noop {} {\bibfield
  {journal} {\bibinfo  {journal} {Nature}\ }\textbf {\bibinfo {volume} {426}},\
  \bibinfo {pages} {55} (\bibinfo {year} {2003})}\BibitemShut {NoStop}%
\bibitem [{\citenamefont {Fiebig}(2005)}]{Fiebig2005}%
  \BibitemOpen
  \bibfield  {author} {\bibinfo {author} {\bibfnamefont {M.}~\bibnamefont
  {Fiebig}},\ }\href@noop {} {\bibfield  {journal} {\bibinfo  {journal} {J.
  Phys. D}\ }\textbf {\bibinfo {volume} {38}},\ \bibinfo {pages} {R123}
  (\bibinfo {year} {2005})}\BibitemShut {NoStop}%
\bibitem [{\citenamefont {Cheong}\ and\ \citenamefont
  {Mostovoy}(2007)}]{cheong2007multiferroics}%
  \BibitemOpen
  \bibfield  {author} {\bibinfo {author} {\bibfnamefont {S.-W.}\ \bibnamefont
  {Cheong}}\ and\ \bibinfo {author} {\bibfnamefont {M.}~\bibnamefont
  {Mostovoy}},\ }\href@noop {} {\bibfield  {journal} {\bibinfo  {journal} {Nat.
  Mater.}\ }\textbf {\bibinfo {volume} {6}},\ \bibinfo {pages} {13} (\bibinfo
  {year} {2007})}\BibitemShut {NoStop}%
\bibitem [{\citenamefont {Arima}(2008)}]{arima2008magneto}%
  \BibitemOpen
  \bibfield  {author} {\bibinfo {author} {\bibfnamefont {T.}~\bibnamefont
  {Arima}},\ }\href@noop {} {\bibfield  {journal} {\bibinfo  {journal} {J.
  Phys.: Condens. Matter}\ }\textbf {\bibinfo {volume} {20}},\ \bibinfo {pages}
  {434211} (\bibinfo {year} {2008})}\BibitemShut {NoStop}%
\bibitem [{\citenamefont {Spaldin}\ \emph {et~al.}(2008)\citenamefont
  {Spaldin}, \citenamefont {Fiebig},\ and\ \citenamefont
  {Mostovoy}}]{Spaldin2008}%
  \BibitemOpen
  \bibfield  {author} {\bibinfo {author} {\bibfnamefont {N.~A.}\ \bibnamefont
  {Spaldin}}, \bibinfo {author} {\bibfnamefont {M.}~\bibnamefont {Fiebig}}, \
  and\ \bibinfo {author} {\bibfnamefont {M.}~\bibnamefont {Mostovoy}},\
  }\href@noop {} {\bibfield  {journal} {\bibinfo  {journal} {J. Phys.: Condens.
  Matter}\ }\textbf {\bibinfo {volume} {20}},\ \bibinfo {pages} {434203}
  (\bibinfo {year} {2008})}\BibitemShut {NoStop}%
\bibitem [{\citenamefont {Mundy}\ \emph {et~al.}(2016)\citenamefont {Mundy},
  \citenamefont {Brooks}, \citenamefont {Holtz}, \citenamefont {Moyer},
  \citenamefont {Das}, \citenamefont {R{\'e}bola}, \citenamefont {Heron},
  \citenamefont {Clarkson}, \citenamefont {Disseler}, \citenamefont {Liu} \emph
  {et~al.}}]{mundy2016atomically}%
  \BibitemOpen
  \bibfield  {author} {\bibinfo {author} {\bibfnamefont {J.~A.}\ \bibnamefont
  {Mundy}}, \bibinfo {author} {\bibfnamefont {C.~M.}\ \bibnamefont {Brooks}},
  \bibinfo {author} {\bibfnamefont {M.~E.}\ \bibnamefont {Holtz}}, \bibinfo
  {author} {\bibfnamefont {J.~A.}\ \bibnamefont {Moyer}}, \bibinfo {author}
  {\bibfnamefont {H.}~\bibnamefont {Das}}, \bibinfo {author} {\bibfnamefont
  {A.~F.}\ \bibnamefont {R{\'e}bola}}, \bibinfo {author} {\bibfnamefont
  {J.~T.}\ \bibnamefont {Heron}}, \bibinfo {author} {\bibfnamefont {J.~D.}\
  \bibnamefont {Clarkson}}, \bibinfo {author} {\bibfnamefont {S.~M.}\
  \bibnamefont {Disseler}}, \bibinfo {author} {\bibfnamefont {Z.}~\bibnamefont
  {Liu}},  \emph {et~al.},\ }\href@noop {} {\bibfield  {journal} {\bibinfo
  {journal} {Nature}\ }\textbf {\bibinfo {volume} {537}},\ \bibinfo {pages}
  {523} (\bibinfo {year} {2016})}\BibitemShut {NoStop}%
\bibitem [{\citenamefont {Baltz}\ \emph {et~al.}(2018)\citenamefont {Baltz},
  \citenamefont {Manchon}, \citenamefont {Tsoi}, \citenamefont {Moriyama},
  \citenamefont {Ono},\ and\ \citenamefont
  {Tserkovnyak}}]{baltz2018antiferromagnetic}%
  \BibitemOpen
  \bibfield  {author} {\bibinfo {author} {\bibfnamefont {V.}~\bibnamefont
  {Baltz}}, \bibinfo {author} {\bibfnamefont {A.}~\bibnamefont {Manchon}},
  \bibinfo {author} {\bibfnamefont {M.}~\bibnamefont {Tsoi}}, \bibinfo {author}
  {\bibfnamefont {T.}~\bibnamefont {Moriyama}}, \bibinfo {author}
  {\bibfnamefont {T.}~\bibnamefont {Ono}}, \ and\ \bibinfo {author}
  {\bibfnamefont {Y.}~\bibnamefont {Tserkovnyak}},\ }\href@noop {} {\bibfield
  {journal} {\bibinfo  {journal} {Rev. Mod. Phys.}\ }\textbf {\bibinfo {volume}
  {90}},\ \bibinfo {pages} {015005} (\bibinfo {year} {2018})}\BibitemShut
  {NoStop}%
\bibitem [{\citenamefont {Jain}\ \emph {et~al.}(2009)\citenamefont {Jain},
  \citenamefont {Ramachandran}, \citenamefont {Clark}, \citenamefont {Zhou},
  \citenamefont {Toby}, \citenamefont {Dalal}, \citenamefont {Kroto},\ and\
  \citenamefont {Cheetham}}]{jain2009multiferroic}%
  \BibitemOpen
  \bibfield  {author} {\bibinfo {author} {\bibfnamefont {P.}~\bibnamefont
  {Jain}}, \bibinfo {author} {\bibfnamefont {V.}~\bibnamefont {Ramachandran}},
  \bibinfo {author} {\bibfnamefont {R.~J.}\ \bibnamefont {Clark}}, \bibinfo
  {author} {\bibfnamefont {H.~D.}\ \bibnamefont {Zhou}}, \bibinfo {author}
  {\bibfnamefont {B.~H.}\ \bibnamefont {Toby}}, \bibinfo {author}
  {\bibfnamefont {N.~S.}\ \bibnamefont {Dalal}}, \bibinfo {author}
  {\bibfnamefont {H.~W.}\ \bibnamefont {Kroto}}, \ and\ \bibinfo {author}
  {\bibfnamefont {A.~K.}\ \bibnamefont {Cheetham}},\ }\href@noop {} {\bibfield
  {journal} {\bibinfo  {journal} {J. Am. Chem. Soc.}\ }\textbf {\bibinfo
  {volume} {131}},\ \bibinfo {pages} {13625} (\bibinfo {year}
  {2009})}\BibitemShut {NoStop}%
\bibitem [{\citenamefont {Xu}\ \emph {et~al.}(2011)\citenamefont {Xu},
  \citenamefont {Zhang}, \citenamefont {Ma}, \citenamefont {Chen},
  \citenamefont {Zhang}, \citenamefont {Cai}, \citenamefont {Wang},
  \citenamefont {Xiong},\ and\ \citenamefont {Gao}}]{xu2011coexistence}%
  \BibitemOpen
  \bibfield  {author} {\bibinfo {author} {\bibfnamefont {G.-C.}\ \bibnamefont
  {Xu}}, \bibinfo {author} {\bibfnamefont {W.}~\bibnamefont {Zhang}}, \bibinfo
  {author} {\bibfnamefont {X.-M.}\ \bibnamefont {Ma}}, \bibinfo {author}
  {\bibfnamefont {Y.-H.}\ \bibnamefont {Chen}}, \bibinfo {author}
  {\bibfnamefont {L.}~\bibnamefont {Zhang}}, \bibinfo {author} {\bibfnamefont
  {H.-L.}\ \bibnamefont {Cai}}, \bibinfo {author} {\bibfnamefont {Z.-M.}\
  \bibnamefont {Wang}}, \bibinfo {author} {\bibfnamefont {R.-G.}\ \bibnamefont
  {Xiong}}, \ and\ \bibinfo {author} {\bibfnamefont {S.}~\bibnamefont {Gao}},\
  }\href@noop {} {\bibfield  {journal} {\bibinfo  {journal} {J. Am. Chem.
  Soc.}\ }\textbf {\bibinfo {volume} {133}},\ \bibinfo {pages} {14948}
  (\bibinfo {year} {2011})}\BibitemShut {NoStop}%
\bibitem [{\citenamefont {Pato-Dold{\'a}n}\ \emph {et~al.}(2013)\citenamefont
  {Pato-Dold{\'a}n}, \citenamefont {G{\'o}mez-Aguirre}, \citenamefont
  {Berm{\'u}dez-Garc{\'\i}a}, \citenamefont {S{\'a}nchez-And{\'u}jar},
  \citenamefont {Fondado}, \citenamefont {Mira}, \citenamefont
  {Castro-Garc{\'\i}a},\ and\ \citenamefont
  {Se{\~n}ar{\'\i}s-Rodr{\'\i}guez}}]{pato2013coexistence}%
  \BibitemOpen
  \bibfield  {author} {\bibinfo {author} {\bibfnamefont {B.}~\bibnamefont
  {Pato-Dold{\'a}n}}, \bibinfo {author} {\bibfnamefont {L.~C.}\ \bibnamefont
  {G{\'o}mez-Aguirre}}, \bibinfo {author} {\bibfnamefont {J.~M.}\ \bibnamefont
  {Berm{\'u}dez-Garc{\'\i}a}}, \bibinfo {author} {\bibfnamefont
  {M.}~\bibnamefont {S{\'a}nchez-And{\'u}jar}}, \bibinfo {author}
  {\bibfnamefont {A.}~\bibnamefont {Fondado}}, \bibinfo {author} {\bibfnamefont
  {J.}~\bibnamefont {Mira}}, \bibinfo {author} {\bibfnamefont {S.}~\bibnamefont
  {Castro-Garc{\'\i}a}}, \ and\ \bibinfo {author} {\bibfnamefont {M.~A.}\
  \bibnamefont {Se{\~n}ar{\'\i}s-Rodr{\'\i}guez}},\ }\href@noop {} {\bibfield
  {journal} {\bibinfo  {journal} {RSC Adv.}\ }\textbf {\bibinfo {volume} {3}},\
  \bibinfo {pages} {22404} (\bibinfo {year} {2013})}\BibitemShut {NoStop}%
\bibitem [{\citenamefont {Ramesh}(2009)}]{ramesh2009materials}%
  \BibitemOpen
  \bibfield  {author} {\bibinfo {author} {\bibfnamefont {R.}~\bibnamefont
  {Ramesh}},\ }\href@noop {} {\bibfield  {journal} {\bibinfo  {journal}
  {Nature}\ }\textbf {\bibinfo {volume} {461}},\ \bibinfo {pages} {1218}
  (\bibinfo {year} {2009})}\BibitemShut {NoStop}%
\bibitem [{\citenamefont {Katsura}\ \emph {et~al.}(2005)\citenamefont
  {Katsura}, \citenamefont {Nagaosa},\ and\ \citenamefont
  {Balatsky}}]{Katsura2005}%
  \BibitemOpen
  \bibfield  {author} {\bibinfo {author} {\bibfnamefont {H.}~\bibnamefont
  {Katsura}}, \bibinfo {author} {\bibfnamefont {N.}~\bibnamefont {Nagaosa}}, \
  and\ \bibinfo {author} {\bibfnamefont {A.~V.}\ \bibnamefont {Balatsky}},\
  }\href {\doibase 10.1103/PhysRevLett.95.057205} {\bibfield  {journal}
  {\bibinfo  {journal} {Phys. Rev. Lett.}\ }\textbf {\bibinfo {volume} {95}},\
  \bibinfo {pages} {057205} (\bibinfo {year} {2005})}\BibitemShut {NoStop}%
\bibitem [{\citenamefont {Sergienko}\ and\ \citenamefont
  {Dagotto}(2006)}]{Sergienko2006}%
  \BibitemOpen
  \bibfield  {author} {\bibinfo {author} {\bibfnamefont {I.~A.}\ \bibnamefont
  {Sergienko}}\ and\ \bibinfo {author} {\bibfnamefont {E.}~\bibnamefont
  {Dagotto}},\ }\href {\doibase 10.1103/PhysRevB.73.094434} {\bibfield
  {journal} {\bibinfo  {journal} {Phys. Rev. B}\ }\textbf {\bibinfo {volume}
  {73}},\ \bibinfo {pages} {094434} (\bibinfo {year} {2006})}\BibitemShut
  {NoStop}%
\bibitem [{\citenamefont {Arima}(2007)}]{Arima2007}%
  \BibitemOpen
  \bibfield  {author} {\bibinfo {author} {\bibfnamefont {T.}~\bibnamefont
  {Arima}},\ }\href@noop {} {\bibfield  {journal} {\bibinfo  {journal} {J.
  Phys. Soc. Jpn.}\ }\textbf {\bibinfo {volume} {76}},\ \bibinfo {pages}
  {073702} (\bibinfo {year} {2007})}\BibitemShut {NoStop}%
\bibitem [{\citenamefont {Sergienko}\ \emph {et~al.}(2006)\citenamefont
  {Sergienko}, \citenamefont {\ifmmode~\mbox{\c{S}}\else \c{S}\fi{}en},\ and\
  \citenamefont {Dagotto}}]{Sergienko2006ferroelectricity}%
  \BibitemOpen
  \bibfield  {author} {\bibinfo {author} {\bibfnamefont {I.~A.}\ \bibnamefont
  {Sergienko}}, \bibinfo {author} {\bibfnamefont {C.}~\bibnamefont
  {\ifmmode~\mbox{\c{S}}\else \c{S}\fi{}en}}, \ and\ \bibinfo {author}
  {\bibfnamefont {E.}~\bibnamefont {Dagotto}},\ }\href {\doibase
  10.1103/PhysRevLett.97.227204} {\bibfield  {journal} {\bibinfo  {journal}
  {Phys. Rev. Lett.}\ }\textbf {\bibinfo {volume} {97}},\ \bibinfo {pages}
  {227204} (\bibinfo {year} {2006})}\BibitemShut {NoStop}%
\bibitem [{\citenamefont {Aoyama}\ \emph {et~al.}(2014)\citenamefont {Aoyama},
  \citenamefont {Yamauchi}, \citenamefont {Iyama}, \citenamefont {Picozzi},
  \citenamefont {Shimizu},\ and\ \citenamefont {Kimura}}]{aoyama2014giant}%
  \BibitemOpen
  \bibfield  {author} {\bibinfo {author} {\bibfnamefont {T.}~\bibnamefont
  {Aoyama}}, \bibinfo {author} {\bibfnamefont {K.}~\bibnamefont {Yamauchi}},
  \bibinfo {author} {\bibfnamefont {A.}~\bibnamefont {Iyama}}, \bibinfo
  {author} {\bibfnamefont {S.}~\bibnamefont {Picozzi}}, \bibinfo {author}
  {\bibfnamefont {K.}~\bibnamefont {Shimizu}}, \ and\ \bibinfo {author}
  {\bibfnamefont {T.}~\bibnamefont {Kimura}},\ }\href@noop {} {\bibfield
  {journal} {\bibinfo  {journal} {Nat. Commun.}\ }\textbf {\bibinfo {volume}
  {5}},\ \bibinfo {pages} {4927} (\bibinfo {year} {2014})}\BibitemShut
  {NoStop}%
\bibitem [{\citenamefont {Aoyama}\ \emph {et~al.}(2015)\citenamefont {Aoyama},
  \citenamefont {Iyama}, \citenamefont {Shimizu},\ and\ \citenamefont
  {Kimura}}]{aoyama2015multiferroicity}%
  \BibitemOpen
  \bibfield  {author} {\bibinfo {author} {\bibfnamefont {T.}~\bibnamefont
  {Aoyama}}, \bibinfo {author} {\bibfnamefont {A.}~\bibnamefont {Iyama}},
  \bibinfo {author} {\bibfnamefont {K.}~\bibnamefont {Shimizu}}, \ and\
  \bibinfo {author} {\bibfnamefont {T.}~\bibnamefont {Kimura}},\ }\href
  {\doibase 10.1103/PhysRevB.91.081107} {\bibfield  {journal} {\bibinfo
  {journal} {Phys. Rev. B}\ }\textbf {\bibinfo {volume} {91}},\ \bibinfo
  {pages} {081107} (\bibinfo {year} {2015})}\BibitemShut {NoStop}%
\bibitem [{\citenamefont {Terada}\ \emph {et~al.}(2016)\citenamefont {Terada},
  \citenamefont {Khalyavin}, \citenamefont {Manuel}, \citenamefont {Osakabe},
  \citenamefont {Kikkawa},\ and\ \citenamefont
  {Kitazawa}}]{terada2016magnetic}%
  \BibitemOpen
  \bibfield  {author} {\bibinfo {author} {\bibfnamefont {N.}~\bibnamefont
  {Terada}}, \bibinfo {author} {\bibfnamefont {D.~D.}\ \bibnamefont
  {Khalyavin}}, \bibinfo {author} {\bibfnamefont {P.}~\bibnamefont {Manuel}},
  \bibinfo {author} {\bibfnamefont {T.}~\bibnamefont {Osakabe}}, \bibinfo
  {author} {\bibfnamefont {A.}~\bibnamefont {Kikkawa}}, \ and\ \bibinfo
  {author} {\bibfnamefont {H.}~\bibnamefont {Kitazawa}},\ }\href {\doibase
  10.1103/PhysRevB.93.081104} {\bibfield  {journal} {\bibinfo  {journal} {Phys.
  Rev. B}\ }\textbf {\bibinfo {volume} {93}},\ \bibinfo {pages} {081104}
  (\bibinfo {year} {2016})}\BibitemShut {NoStop}%
\bibitem [{\citenamefont {Delaney}\ \emph {et~al.}(2009)\citenamefont
  {Delaney}, \citenamefont {Mostovoy},\ and\ \citenamefont
  {Spaldin}}]{Delaney2009}%
  \BibitemOpen
  \bibfield  {author} {\bibinfo {author} {\bibfnamefont {K.~T.}\ \bibnamefont
  {Delaney}}, \bibinfo {author} {\bibfnamefont {M.}~\bibnamefont {Mostovoy}}, \
  and\ \bibinfo {author} {\bibfnamefont {N.~A.}\ \bibnamefont {Spaldin}},\
  }\href {\doibase 10.1103/PhysRevLett.102.157203} {\bibfield  {journal}
  {\bibinfo  {journal} {Phys. Rev. Lett.}\ }\textbf {\bibinfo {volume} {102}},\
  \bibinfo {pages} {157203} (\bibinfo {year} {2009})}\BibitemShut {NoStop}%
\bibitem [{\citenamefont {Giester}\ \emph {et~al.}(2007)\citenamefont
  {Giester}, \citenamefont {Kolitsch}, \citenamefont {Leverett}, \citenamefont
  {Turner},\ and\ \citenamefont {Williams}}]{giester2007crystal}%
  \BibitemOpen
  \bibfield  {author} {\bibinfo {author} {\bibfnamefont {G.}~\bibnamefont
  {Giester}}, \bibinfo {author} {\bibfnamefont {U.}~\bibnamefont {Kolitsch}},
  \bibinfo {author} {\bibfnamefont {P.}~\bibnamefont {Leverett}}, \bibinfo
  {author} {\bibfnamefont {P.}~\bibnamefont {Turner}}, \ and\ \bibinfo {author}
  {\bibfnamefont {P.~A.}\ \bibnamefont {Williams}},\ }\href {\doibase
  10.1127/0935-1221/2007/0019-0075} {\bibfield  {journal} {\bibinfo  {journal}
  {Eur. J. Mineral.}\ }\textbf {\bibinfo {volume} {19}},\ \bibinfo {pages} {75}
  (\bibinfo {year} {2007})}\BibitemShut {NoStop}%
\bibitem [{\citenamefont {Hwu}\ \emph {et~al.}(2002)\citenamefont {Hwu},
  \citenamefont {Ulutagay-Kartin}, \citenamefont {Clayhold}, \citenamefont
  {Mackay}, \citenamefont {Wardojo}, \citenamefont {O'Connor},\ and\
  \citenamefont {Krawiec}}]{Hwu2002}%
  \BibitemOpen
  \bibfield  {author} {\bibinfo {author} {\bibfnamefont {S.-J.}\ \bibnamefont
  {Hwu}}, \bibinfo {author} {\bibfnamefont {M.}~\bibnamefont
  {Ulutagay-Kartin}}, \bibinfo {author} {\bibfnamefont {J.~A.}\ \bibnamefont
  {Clayhold}}, \bibinfo {author} {\bibfnamefont {R.}~\bibnamefont {Mackay}},
  \bibinfo {author} {\bibfnamefont {T.~A.}\ \bibnamefont {Wardojo}}, \bibinfo
  {author} {\bibfnamefont {C.~J.}\ \bibnamefont {O'Connor}}, \ and\ \bibinfo
  {author} {\bibfnamefont {M.}~\bibnamefont {Krawiec}},\ }\href@noop {}
  {\bibfield  {journal} {\bibinfo  {journal} {J. Am. Chem. Soc.}\ }\textbf
  {\bibinfo {volume} {124}},\ \bibinfo {pages} {12404} (\bibinfo {year}
  {2002})}\BibitemShut {NoStop}%
\bibitem [{\citenamefont {Williams}\ \emph {et~al.}(2015)\citenamefont
  {Williams}, \citenamefont {Marshall},\ and\ \citenamefont
  {Weller}}]{Williams2015}%
  \BibitemOpen
  \bibfield  {author} {\bibinfo {author} {\bibfnamefont {E.~R.}\ \bibnamefont
  {Williams}}, \bibinfo {author} {\bibfnamefont {K.}~\bibnamefont {Marshall}},
  \ and\ \bibinfo {author} {\bibfnamefont {M.~T.}\ \bibnamefont {Weller}},\
  }\href@noop {} {\bibfield  {journal} {\bibinfo  {journal} {CrystEngComm}\
  }\textbf {\bibinfo {volume} {17}},\ \bibinfo {pages} {160} (\bibinfo {year}
  {2015})}\BibitemShut {NoStop}%
\bibitem [{\citenamefont {Kimura}\ \emph
  {et~al.}(2016{\natexlab{a}})\citenamefont {Kimura}, \citenamefont {Sera},\
  and\ \citenamefont {Kimura}}]{KKimura2016}%
  \BibitemOpen
  \bibfield  {author} {\bibinfo {author} {\bibfnamefont {K.}~\bibnamefont
  {Kimura}}, \bibinfo {author} {\bibfnamefont {M.}~\bibnamefont {Sera}}, \ and\
  \bibinfo {author} {\bibfnamefont {T.}~\bibnamefont {Kimura}},\ }\href
  {\doibase 10.1021/acs.inorgchem.5b02622} {\bibfield  {journal} {\bibinfo
  {journal} {Inorg. Chem.}\ }\textbf {\bibinfo {volume} {55}},\ \bibinfo
  {pages} {1002} (\bibinfo {year} {2016}{\natexlab{a}})}\BibitemShut {NoStop}%
\bibitem [{\citenamefont {Kimura}\ \emph
  {et~al.}(2016{\natexlab{b}})\citenamefont {Kimura}, \citenamefont
  {Babkevich}, \citenamefont {Sera}, \citenamefont {Toyoda}, \citenamefont
  {Yamauchi}, \citenamefont {Tucker}, \citenamefont {Martius}, \citenamefont
  {Fennell}, \citenamefont {Manuel}, \citenamefont {Khalyavin}, \citenamefont
  {Johnson}, \citenamefont {Nakano}, \citenamefont {Nozue}, \citenamefont
  {{R$\o{}$nnow}},\ and\ \citenamefont {Kimura}}]{KKimura2016b}%
  \BibitemOpen
  \bibfield  {author} {\bibinfo {author} {\bibfnamefont {K.}~\bibnamefont
  {Kimura}}, \bibinfo {author} {\bibfnamefont {P.}~\bibnamefont {Babkevich}},
  \bibinfo {author} {\bibfnamefont {M.}~\bibnamefont {Sera}}, \bibinfo {author}
  {\bibfnamefont {M.}~\bibnamefont {Toyoda}}, \bibinfo {author} {\bibfnamefont
  {K.}~\bibnamefont {Yamauchi}}, \bibinfo {author} {\bibfnamefont {G.~S.}\
  \bibnamefont {Tucker}}, \bibinfo {author} {\bibfnamefont {J.}~\bibnamefont
  {Martius}}, \bibinfo {author} {\bibfnamefont {T.}~\bibnamefont {Fennell}},
  \bibinfo {author} {\bibfnamefont {P.}~\bibnamefont {Manuel}}, \bibinfo
  {author} {\bibfnamefont {D.~D.}\ \bibnamefont {Khalyavin}}, \bibinfo {author}
  {\bibfnamefont {R.~D.}\ \bibnamefont {Johnson}}, \bibinfo {author}
  {\bibfnamefont {T.}~\bibnamefont {Nakano}}, \bibinfo {author} {\bibfnamefont
  {Y.}~\bibnamefont {Nozue}}, \bibinfo {author} {\bibfnamefont {H.~M.}\
  \bibnamefont {{R$\o{}$nnow}}}, \ and\ \bibinfo {author} {\bibfnamefont
  {T.}~\bibnamefont {Kimura}},\ }\href@noop {} {\bibfield  {journal} {\bibinfo
  {journal} {Nat. Commun.}\ }\textbf {\bibinfo {volume} {7}},\ \bibinfo {pages}
  {13039} (\bibinfo {year} {2016}{\natexlab{b}})}\BibitemShut {NoStop}%
\bibitem [{\citenamefont {Kimura}\ \emph {et~al.}(2018)\citenamefont {Kimura},
  \citenamefont {Toyoda}, \citenamefont {Babkevich}, \citenamefont {Yamauchi},
  \citenamefont {Sera}, \citenamefont {Nassif}, \citenamefont {R\o{}nnow},\
  and\ \citenamefont {Kimura}}]{kimura2018a}%
  \BibitemOpen
  \bibfield  {author} {\bibinfo {author} {\bibfnamefont {K.}~\bibnamefont
  {Kimura}}, \bibinfo {author} {\bibfnamefont {M.}~\bibnamefont {Toyoda}},
  \bibinfo {author} {\bibfnamefont {P.}~\bibnamefont {Babkevich}}, \bibinfo
  {author} {\bibfnamefont {K.}~\bibnamefont {Yamauchi}}, \bibinfo {author}
  {\bibfnamefont {M.}~\bibnamefont {Sera}}, \bibinfo {author} {\bibfnamefont
  {V.}~\bibnamefont {Nassif}}, \bibinfo {author} {\bibfnamefont {H.~M.}\
  \bibnamefont {R\o{}nnow}}, \ and\ \bibinfo {author} {\bibfnamefont
  {T.}~\bibnamefont {Kimura}},\ }\href {\doibase 10.1103/PhysRevB.97.134418}
  {\bibfield  {journal} {\bibinfo  {journal} {Phys. Rev. B}\ }\textbf {\bibinfo
  {volume} {97}},\ \bibinfo {pages} {134418} (\bibinfo {year}
  {2018})}\BibitemShut {NoStop}%
\bibitem [{\citenamefont {Kato}\ \emph {et~al.}(2017)\citenamefont {Kato},
  \citenamefont {Kimura}, \citenamefont {Miyake}, \citenamefont {Tokunaga},
  \citenamefont {Matsuo}, \citenamefont {Kindo}, \citenamefont {Akaki},
  \citenamefont {Hagiwara}, \citenamefont {Sera}, \citenamefont {Kimura},\ and\
  \citenamefont {Motome}}]{Kato2017}%
  \BibitemOpen
  \bibfield  {author} {\bibinfo {author} {\bibfnamefont {Y.}~\bibnamefont
  {Kato}}, \bibinfo {author} {\bibfnamefont {K.}~\bibnamefont {Kimura}},
  \bibinfo {author} {\bibfnamefont {A.}~\bibnamefont {Miyake}}, \bibinfo
  {author} {\bibfnamefont {M.}~\bibnamefont {Tokunaga}}, \bibinfo {author}
  {\bibfnamefont {A.}~\bibnamefont {Matsuo}}, \bibinfo {author} {\bibfnamefont
  {K.}~\bibnamefont {Kindo}}, \bibinfo {author} {\bibfnamefont
  {M.}~\bibnamefont {Akaki}}, \bibinfo {author} {\bibfnamefont
  {M.}~\bibnamefont {Hagiwara}}, \bibinfo {author} {\bibfnamefont
  {M.}~\bibnamefont {Sera}}, \bibinfo {author} {\bibfnamefont {T.}~\bibnamefont
  {Kimura}}, \ and\ \bibinfo {author} {\bibfnamefont {Y.}~\bibnamefont
  {Motome}},\ }\href {\doibase 10.1103/PhysRevLett.118.107601} {\bibfield
  {journal} {\bibinfo  {journal} {Phys. Rev. Lett.}\ }\textbf {\bibinfo
  {volume} {118}},\ \bibinfo {pages} {107601} (\bibinfo {year}
  {2017})}\BibitemShut {NoStop}%
\bibitem [{\citenamefont {Mitamura}\ \emph {et~al.}(2007)\citenamefont
  {Mitamura}, \citenamefont {Mitsuda}, \citenamefont {Kanetsuki}, \citenamefont
  {Aruga~Katori}, \citenamefont {Sakakibara},\ and\ \citenamefont
  {Kindo}}]{mitamura2007dielectric}%
  \BibitemOpen
  \bibfield  {author} {\bibinfo {author} {\bibfnamefont {H.}~\bibnamefont
  {Mitamura}}, \bibinfo {author} {\bibfnamefont {S.}~\bibnamefont {Mitsuda}},
  \bibinfo {author} {\bibfnamefont {S.}~\bibnamefont {Kanetsuki}}, \bibinfo
  {author} {\bibfnamefont {H.}~\bibnamefont {Aruga~Katori}}, \bibinfo {author}
  {\bibfnamefont {T.}~\bibnamefont {Sakakibara}}, \ and\ \bibinfo {author}
  {\bibfnamefont {K.}~\bibnamefont {Kindo}},\ }\href@noop {} {\bibfield
  {journal} {\bibinfo  {journal} {J. Phys. Soc. Jpn.}\ }\textbf {\bibinfo
  {volume} {76}},\ \bibinfo {pages} {094709} (\bibinfo {year}
  {2007})}\BibitemShut {NoStop}%
\bibitem [{\citenamefont {Momma}\ and\ \citenamefont
  {Izumi}(2011)}]{Momma2011}%
  \BibitemOpen
  \bibfield  {author} {\bibinfo {author} {\bibfnamefont {K.}~\bibnamefont
  {Momma}}\ and\ \bibinfo {author} {\bibfnamefont {F.}~\bibnamefont {Izumi}},\
  }\href@noop {} {\bibfield  {journal} {\bibinfo  {journal} {J. Appl.
  Crystallogr.}\ }\textbf {\bibinfo {volume} {44}},\ \bibinfo {pages} {1272}
  (\bibinfo {year} {2011})}\BibitemShut {NoStop}%
\bibitem [{\citenamefont {Kresse}\ and\ \citenamefont
  {Furthm\"uller}(1996)}]{Kresse1996}%
  \BibitemOpen
  \bibfield  {author} {\bibinfo {author} {\bibfnamefont {G.}~\bibnamefont
  {Kresse}}\ and\ \bibinfo {author} {\bibfnamefont {J.}~\bibnamefont
  {Furthm\"uller}},\ }\href@noop {} {\bibfield  {journal} {\bibinfo  {journal}
  {Phys. Rev. B}\ }\textbf {\bibinfo {volume} {54}},\ \bibinfo {pages} {11169}
  (\bibinfo {year} {1996})}\BibitemShut {NoStop}%
\bibitem [{\citenamefont {Perdew}\ \emph {et~al.}(1996)\citenamefont {Perdew},
  \citenamefont {Burke},\ and\ \citenamefont {Ernzerhof}}]{PBE1996}%
  \BibitemOpen
  \bibfield  {author} {\bibinfo {author} {\bibfnamefont {J.~P.}\ \bibnamefont
  {Perdew}}, \bibinfo {author} {\bibfnamefont {K.}~\bibnamefont {Burke}}, \
  and\ \bibinfo {author} {\bibfnamefont {M.}~\bibnamefont {Ernzerhof}},\
  }\href@noop {} {\bibfield  {journal} {\bibinfo  {journal} {Phys. Rev. Lett.}\
  }\textbf {\bibinfo {volume} {77}},\ \bibinfo {pages} {3865} (\bibinfo {year}
  {1996})}\BibitemShut {NoStop}%
\bibitem [{\citenamefont {Liechtenstein}\ \emph {et~al.}(1995)\citenamefont
  {Liechtenstein}, \citenamefont {Anisimov},\ and\ \citenamefont
  {Zaanen}}]{Liechtenstein1995}%
  \BibitemOpen
  \bibfield  {author} {\bibinfo {author} {\bibfnamefont {A.~I.}\ \bibnamefont
  {Liechtenstein}}, \bibinfo {author} {\bibfnamefont {V.~I.}\ \bibnamefont
  {Anisimov}}, \ and\ \bibinfo {author} {\bibfnamefont {J.}~\bibnamefont
  {Zaanen}},\ }\href@noop {} {\bibfield  {journal} {\bibinfo  {journal} {Phys.
  Rev. B}\ }\textbf {\bibinfo {volume} {52}},\ \bibinfo {pages} {R5467}
  (\bibinfo {year} {1995})}\BibitemShut {NoStop}%
\bibitem [{\citenamefont {King-Smith}\ and\ \citenamefont
  {Vanderbilt}(1993)}]{smith1993theory}%
  \BibitemOpen
  \bibfield  {author} {\bibinfo {author} {\bibfnamefont {R.~D.}\ \bibnamefont
  {King-Smith}}\ and\ \bibinfo {author} {\bibfnamefont {D.}~\bibnamefont
  {Vanderbilt}},\ }\href {\doibase 10.1103/PhysRevB.47.1651} {\bibfield
  {journal} {\bibinfo  {journal} {Phys. Rev. B}\ }\textbf {\bibinfo {volume}
  {47}},\ \bibinfo {pages} {1651} (\bibinfo {year} {1993})}\BibitemShut
  {NoStop}%
\bibitem [{\citenamefont {Resta}(1994)}]{Resta1994macroscopic}%
  \BibitemOpen
  \bibfield  {author} {\bibinfo {author} {\bibfnamefont {R.}~\bibnamefont
  {Resta}},\ }\href {\doibase 10.1103/RevModPhys.66.899} {\bibfield  {journal}
  {\bibinfo  {journal} {Rev. Mod. Phys.}\ }\textbf {\bibinfo {volume} {66}},\
  \bibinfo {pages} {899} (\bibinfo {year} {1994})}\BibitemShut {NoStop}%
\bibitem [{\citenamefont {Hur}\ \emph {et~al.}(2004{\natexlab{a}})\citenamefont
  {Hur}, \citenamefont {Park}, \citenamefont {Sharma}, \citenamefont {Guha},\
  and\ \citenamefont {Cheong}}]{hur2004colossal}%
  \BibitemOpen
  \bibfield  {author} {\bibinfo {author} {\bibfnamefont {N.}~\bibnamefont
  {Hur}}, \bibinfo {author} {\bibfnamefont {S.}~\bibnamefont {Park}}, \bibinfo
  {author} {\bibfnamefont {P.~A.}\ \bibnamefont {Sharma}}, \bibinfo {author}
  {\bibfnamefont {S.}~\bibnamefont {Guha}}, \ and\ \bibinfo {author}
  {\bibfnamefont {S.-W.}\ \bibnamefont {Cheong}},\ }\href {\doibase
  10.1103/PhysRevLett.93.107207} {\bibfield  {journal} {\bibinfo  {journal}
  {Phys. Rev. Lett.}\ }\textbf {\bibinfo {volume} {93}},\ \bibinfo {pages}
  {107207} (\bibinfo {year} {2004}{\natexlab{a}})}\BibitemShut {NoStop}%
\bibitem [{\citenamefont {Goto}\ \emph {et~al.}(2004)\citenamefont {Goto},
  \citenamefont {Kimura}, \citenamefont {Lawes}, \citenamefont {Ramirez},\ and\
  \citenamefont {Tokura}}]{goto2004ferroelectricity}%
  \BibitemOpen
  \bibfield  {author} {\bibinfo {author} {\bibfnamefont {T.}~\bibnamefont
  {Goto}}, \bibinfo {author} {\bibfnamefont {T.}~\bibnamefont {Kimura}},
  \bibinfo {author} {\bibfnamefont {G.}~\bibnamefont {Lawes}}, \bibinfo
  {author} {\bibfnamefont {A.~P.}\ \bibnamefont {Ramirez}}, \ and\ \bibinfo
  {author} {\bibfnamefont {Y.}~\bibnamefont {Tokura}},\ }\href {\doibase
  10.1103/PhysRevLett.92.257201} {\bibfield  {journal} {\bibinfo  {journal}
  {Phys. Rev. Lett.}\ }\textbf {\bibinfo {volume} {92}},\ \bibinfo {pages}
  {257201} (\bibinfo {year} {2004})}\BibitemShut {NoStop}%
\bibitem [{\citenamefont {Warren}\ \emph {et~al.}(1995)\citenamefont {Warren},
  \citenamefont {Dimos}, \citenamefont {Pike}, \citenamefont {Tuttle},
  \citenamefont {Raymond}, \citenamefont {Ramesh},\ and\ \citenamefont
  {Evans~Jr}}]{warren1995voltage}%
  \BibitemOpen
  \bibfield  {author} {\bibinfo {author} {\bibfnamefont {W.}~\bibnamefont
  {Warren}}, \bibinfo {author} {\bibfnamefont {D.}~\bibnamefont {Dimos}},
  \bibinfo {author} {\bibfnamefont {G.}~\bibnamefont {Pike}}, \bibinfo {author}
  {\bibfnamefont {B.}~\bibnamefont {Tuttle}}, \bibinfo {author} {\bibfnamefont
  {M.}~\bibnamefont {Raymond}}, \bibinfo {author} {\bibfnamefont
  {R.}~\bibnamefont {Ramesh}}, \ and\ \bibinfo {author} {\bibfnamefont
  {J.}~\bibnamefont {Evans~Jr}},\ }\href@noop {} {\bibfield  {journal}
  {\bibinfo  {journal} {Appl. Phys. Lett.}\ }\textbf {\bibinfo {volume} {67}},\
  \bibinfo {pages} {866} (\bibinfo {year} {1995})}\BibitemShut {NoStop}%
\bibitem [{\citenamefont {Moriya}(1960)}]{moriya1960anisotropic}%
  \BibitemOpen
  \bibfield  {author} {\bibinfo {author} {\bibfnamefont {T.}~\bibnamefont
  {Moriya}},\ }\href {\doibase 10.1103/PhysRev.120.91} {\bibfield  {journal}
  {\bibinfo  {journal} {Phys. Rev.}\ }\textbf {\bibinfo {volume} {120}},\
  \bibinfo {pages} {91} (\bibinfo {year} {1960})}\BibitemShut {NoStop}%
\bibitem [{\citenamefont {Kimura}(2012)}]{kimura2012magnetoelectric}%
  \BibitemOpen
  \bibfield  {author} {\bibinfo {author} {\bibfnamefont {T.}~\bibnamefont
  {Kimura}},\ }\href@noop {} {\bibfield  {journal} {\bibinfo  {journal} {Annu.
  Rev. Condens. Matter Phys.}\ }\textbf {\bibinfo {volume} {3}},\ \bibinfo
  {pages} {93} (\bibinfo {year} {2012})}\BibitemShut {NoStop}%
\bibitem [{\citenamefont {Tokunaga}\ \emph {et~al.}(2008)\citenamefont
  {Tokunaga}, \citenamefont {Iguchi}, \citenamefont {Arima},\ and\
  \citenamefont {Tokura}}]{tokunaga2008magnetic}%
  \BibitemOpen
  \bibfield  {author} {\bibinfo {author} {\bibfnamefont {Y.}~\bibnamefont
  {Tokunaga}}, \bibinfo {author} {\bibfnamefont {S.}~\bibnamefont {Iguchi}},
  \bibinfo {author} {\bibfnamefont {T.}~\bibnamefont {Arima}}, \ and\ \bibinfo
  {author} {\bibfnamefont {Y.}~\bibnamefont {Tokura}},\ }\href {\doibase
  10.1103/PhysRevLett.101.097205} {\bibfield  {journal} {\bibinfo  {journal}
  {Phys. Rev. Lett.}\ }\textbf {\bibinfo {volume} {101}},\ \bibinfo {pages}
  {097205} (\bibinfo {year} {2008})}\BibitemShut {NoStop}%
\bibitem [{\citenamefont {Ishiwata}\ \emph {et~al.}(2011)\citenamefont
  {Ishiwata}, \citenamefont {Tokunaga}, \citenamefont {Taguchi},\ and\
  \citenamefont {Tokura}}]{ishiwata2011high}%
  \BibitemOpen
  \bibfield  {author} {\bibinfo {author} {\bibfnamefont {S.}~\bibnamefont
  {Ishiwata}}, \bibinfo {author} {\bibfnamefont {Y.}~\bibnamefont {Tokunaga}},
  \bibinfo {author} {\bibfnamefont {Y.}~\bibnamefont {Taguchi}}, \ and\
  \bibinfo {author} {\bibfnamefont {Y.}~\bibnamefont {Tokura}},\ }\href@noop {}
  {\bibfield  {journal} {\bibinfo  {journal} {J. Am. Chem. Soc.}\ }\textbf
  {\bibinfo {volume} {133}},\ \bibinfo {pages} {13818} (\bibinfo {year}
  {2011})}\BibitemShut {NoStop}%
\bibitem [{\citenamefont {Picozzi}\ \emph {et~al.}(2007)\citenamefont
  {Picozzi}, \citenamefont {Yamauchi}, \citenamefont {Sanyal}, \citenamefont
  {Sergienko},\ and\ \citenamefont {Dagotto}}]{picozzi2007dual}%
  \BibitemOpen
  \bibfield  {author} {\bibinfo {author} {\bibfnamefont {S.}~\bibnamefont
  {Picozzi}}, \bibinfo {author} {\bibfnamefont {K.}~\bibnamefont {Yamauchi}},
  \bibinfo {author} {\bibfnamefont {B.}~\bibnamefont {Sanyal}}, \bibinfo
  {author} {\bibfnamefont {I.~A.}\ \bibnamefont {Sergienko}}, \ and\ \bibinfo
  {author} {\bibfnamefont {E.}~\bibnamefont {Dagotto}},\ }\href {\doibase
  10.1103/PhysRevLett.99.227201} {\bibfield  {journal} {\bibinfo  {journal}
  {Phys. Rev. Lett.}\ }\textbf {\bibinfo {volume} {99}},\ \bibinfo {pages}
  {227201} (\bibinfo {year} {2007})}\BibitemShut {NoStop}%
\bibitem [{\citenamefont {Murakawa}\ \emph {et~al.}(2012)\citenamefont
  {Murakawa}, \citenamefont {Onose}, \citenamefont {Miyahara}, \citenamefont
  {Furukawa},\ and\ \citenamefont {Tokura}}]{murakawa2012comprehensive}%
  \BibitemOpen
  \bibfield  {author} {\bibinfo {author} {\bibfnamefont {H.}~\bibnamefont
  {Murakawa}}, \bibinfo {author} {\bibfnamefont {Y.}~\bibnamefont {Onose}},
  \bibinfo {author} {\bibfnamefont {S.}~\bibnamefont {Miyahara}}, \bibinfo
  {author} {\bibfnamefont {N.}~\bibnamefont {Furukawa}}, \ and\ \bibinfo
  {author} {\bibfnamefont {Y.}~\bibnamefont {Tokura}},\ }\href {\doibase
  10.1103/PhysRevB.85.174106} {\bibfield  {journal} {\bibinfo  {journal} {Phys.
  Rev. B}\ }\textbf {\bibinfo {volume} {85}},\ \bibinfo {pages} {174106}
  (\bibinfo {year} {2012})}\BibitemShut {NoStop}%
\bibitem [{\citenamefont {Akaki}\ \emph {et~al.}(2012)\citenamefont {Akaki},
  \citenamefont {Iwamoto}, \citenamefont {Kihara}, \citenamefont {Tokunaga},\
  and\ \citenamefont {Kuwahara}}]{akaki2012multiferroic}%
  \BibitemOpen
  \bibfield  {author} {\bibinfo {author} {\bibfnamefont {M.}~\bibnamefont
  {Akaki}}, \bibinfo {author} {\bibfnamefont {H.}~\bibnamefont {Iwamoto}},
  \bibinfo {author} {\bibfnamefont {T.}~\bibnamefont {Kihara}}, \bibinfo
  {author} {\bibfnamefont {M.}~\bibnamefont {Tokunaga}}, \ and\ \bibinfo
  {author} {\bibfnamefont {H.}~\bibnamefont {Kuwahara}},\ }\href {\doibase
  10.1103/PhysRevB.86.060413} {\bibfield  {journal} {\bibinfo  {journal} {Phys.
  Rev. B}\ }\textbf {\bibinfo {volume} {86}},\ \bibinfo {pages} {060413}
  (\bibinfo {year} {2012})}\BibitemShut {NoStop}%
\bibitem [{\citenamefont {Hur}\ \emph {et~al.}(2004{\natexlab{b}})\citenamefont
  {Hur}, \citenamefont {Park}, \citenamefont {Sharma}, \citenamefont {Ahn},
  \citenamefont {Guha},\ and\ \citenamefont {Cheong}}]{hur2004electric}%
  \BibitemOpen
  \bibfield  {author} {\bibinfo {author} {\bibfnamefont {N.}~\bibnamefont
  {Hur}}, \bibinfo {author} {\bibfnamefont {S.}~\bibnamefont {Park}}, \bibinfo
  {author} {\bibfnamefont {P.}~\bibnamefont {Sharma}}, \bibinfo {author}
  {\bibfnamefont {J.}~\bibnamefont {Ahn}}, \bibinfo {author} {\bibfnamefont
  {S.}~\bibnamefont {Guha}}, \ and\ \bibinfo {author} {\bibfnamefont
  {S.}~\bibnamefont {Cheong}},\ }\href@noop {} {\bibfield  {journal} {\bibinfo
  {journal} {Nature}\ }\textbf {\bibinfo {volume} {429}},\ \bibinfo {pages}
  {392} (\bibinfo {year} {2004}{\natexlab{b}})}\BibitemShut {NoStop}%
\bibitem [{\citenamefont {Kim}\ \emph {et~al.}(2009)\citenamefont {Kim},
  \citenamefont {Haam}, \citenamefont {Oh}, \citenamefont {Park}, \citenamefont
  {Cheong}, \citenamefont {Sharma}, \citenamefont {Jaime}, \citenamefont
  {Harrison}, \citenamefont {Han}, \citenamefont {Jeon}, \citenamefont
  {Coleman},\ and\ \citenamefont {Kim}}]{kim2009observation}%
  \BibitemOpen
  \bibfield  {author} {\bibinfo {author} {\bibfnamefont {J.~W.}\ \bibnamefont
  {Kim}}, \bibinfo {author} {\bibfnamefont {S.~Y.}\ \bibnamefont {Haam}},
  \bibinfo {author} {\bibfnamefont {Y.~S.}\ \bibnamefont {Oh}}, \bibinfo
  {author} {\bibfnamefont {S.}~\bibnamefont {Park}}, \bibinfo {author}
  {\bibfnamefont {S.-W.}\ \bibnamefont {Cheong}}, \bibinfo {author}
  {\bibfnamefont {P.~A.}\ \bibnamefont {Sharma}}, \bibinfo {author}
  {\bibfnamefont {M.}~\bibnamefont {Jaime}}, \bibinfo {author} {\bibfnamefont
  {N.}~\bibnamefont {Harrison}}, \bibinfo {author} {\bibfnamefont {J.~H.}\
  \bibnamefont {Han}}, \bibinfo {author} {\bibfnamefont {G.-S.}\ \bibnamefont
  {Jeon}}, \bibinfo {author} {\bibfnamefont {P.}~\bibnamefont {Coleman}}, \
  and\ \bibinfo {author} {\bibfnamefont {K.~H.}\ \bibnamefont {Kim}},\
  }\href@noop {} {\bibfield  {journal} {\bibinfo  {journal} {Proc. Natl. Acad.
  Sci. U.S.A.}\ }\textbf {\bibinfo {volume} {106}},\ \bibinfo {pages} {15573}
  (\bibinfo {year} {2009})}\BibitemShut {NoStop}%
\end{thebibliography}
\end{document}